\documentclass{article}

\usepackage{microtype}
\usepackage{graphicx}
\usepackage{subfigure}
\usepackage{booktabs} 

\usepackage{url}
\usepackage{amssymb,amsmath,enumitem,url}
\usepackage{graphicx}
\usepackage{changepage}
\usepackage{stmaryrd}
\usepackage{mathrsfs}
\usepackage{rsfso}
\usepackage{listings}
\usepackage{xcolor}
\usepackage{mathtools}
\usepackage{caption}
\usepackage{subcaption}
\usepackage{float}
\usepackage{hyperref}
\usepackage[capitalize]{cleveref}
\usepackage{array}
\usepackage{stmaryrd}
\usepackage{stfloats}

\usepackage{xspace}

\crefname{figure}{Fig.}{Figs.}
\crefformat{figure}{#2Fig.~#1#3}
\crefrangeformat{figure}{#3Figs.~#1--#2#4}

\Crefname{figure}{Fig.}{Figs.}
\Crefformat{figure}{#2Fig.~#1#3}
\Crefrangeformat{figure}{#3Figs.~#1--#2#4}

\newcommand{\figline}{\par\noindent\rule{\columnwidth}{0.4pt}\par\vspace{0.5ex}}
\newcommand{\doublefigline}{\par\noindent\rule{\textwidth}{0.4pt}\par}
\newcommand{\pageonecrunch}{\vspace{-5pt}}

\crefname{equation}{Eq.}{Eqs.}
\crefformat{equation}{#2Eq.~#1#3}
\crefrangeformat{equation}{#3Eqs.~#1--#2#4}

\Crefname{equation}{Eq.}{Eqs.}
\Crefformat{equation}{#2Eq.~#1#3}
\Crefrangeformat{equation}{#3Eqs.~#1--#2#4}

\newcommand{\define}[1]{{\bf #1}}
\newcommand{\definep}[1]{(\define{#1})}

\newcommand{\data}[2]{x_{#1}^{#2}} 
\newcommand{\genstate}[2]{g_{#1}^{#2}} 
\newcommand{\infinput}[2]{u_{#1}^{#2}} 
\newcommand{\msg}[3]{m_{#1}^{#2 \rightarrow #3}} 
\newcommand{\factors}[2]{f_{#1}^{#2}} 
\newcommand{\rates}[2]{r_{#1}^{#2}} 

\newcommand{\stim}[2]{s_{#1}^{#2}} 
\newcommand{\decision}[1]{d_{#1}} 

\newcommand{\encstate}[2]{e_{#1}^{#2}}
\newcommand{\encfwdstate}[2]{e_{#1}^{#2,+}}
\newcommand{\encbwdstate}[2]{e_{#1}^{#2,-}}
\newcommand{\constate}[2]{c_{#1}^{#2}}

\newcommand{\klweight}[1]{\beta_{#1}}
\newcommand{\Ltwoweight}{\alpha} 
\newcommand{\ltwoweight}{\alpha} 

\newcommand{\gru}[2]{\text{GRU}_{#1}^{#2}} 

\newcommand{\encrnn}[2]{\gru{\text{enc},#1}{#2}} 
\newcommand{\encrnng}[1]{\encrnn{\genstate{}{}}{#1}}
\newcommand{\encbwdrnng}[1]{\encrnng{#1,-}}
\newcommand{\encfwdrnng}[1]{\encrnng{#1,+}}

\newcommand{\conrnn}[1]{\gru{\text{con}}{#1}} 
\newcommand{\genrnn}[1]{\gru{\text{gen}}{#1}} 

\newcommand{\scos}{S_\text{cos}}

\newcommand{\mrlfads}{\mbox{MR-LFADS}\xspace}
\newcommand{\mrlfadsvariant}[1]{\mbox{MR-LFADS(#1)}\xspace}
\newcommand{\mrlfadsr}{\mrlfadsvariant{R}}
\newcommand{\mrlfadss}{\mrlfadsvariant{S}}
\newcommand{\mrlfadsf}{\mrlfadsvariant{F}}
\newcommand{\mrlfadsg}{\mrlfadsvariant{G}}
\newcommand{\mrlfadsrfg}{\mrlfadsr, (F), and (G)}

\newcommand{\mrsds}{\mbox{MR-SDS}\xspace}
\newcommand{\mpsrslds}{\mbox{mp-srSLDS}\xspace}

\usepackage[accepted]{icml2025}

\icmltitlerunning{Accurate Identification of Communication Between Multiple Interacting Neural Populations}

\begin{document}

\twocolumn[
\icmltitle{Accurate Identification of Communication \\ Between Multiple Interacting Neural Populations}



\icmlsetsymbol{equal}{*}

\begin{icmlauthorlist}
\icmlauthor{Belle Liu}{gpn}
\icmlauthor{Jacob Sacks}{cse}
\icmlauthor{Matthew D. Golub}{cse}
\end{icmlauthorlist}

\icmlaffiliation{cse}{Paul G. Allen School of Computer Science \& Engineering, University of Washington}
\icmlaffiliation{gpn}{Graduate Program in Neuroscience, University of Washington;}

\icmlcorrespondingauthor{Matthew Golub}{mgolub@cs.washington.edu}

\icmlkeywords{Computational Neuroscience, NeuroAI, Deep Learning for Neuroscience, Neural Population Dynamics, Brain-Wide Communication, Multi-Region Brain Modeling}

\vskip 0.28in
]



\printAffiliationsAndNotice{}  

\begin{abstract}
Neural recording technologies now enable simultaneous recording of population activity across many brain regions, motivating the development of data-driven models of inter-regional communication. However, existing models can struggle to disentangle the influences that drive recorded population activity, leading to inaccurate portraits of communication. Here, we introduce Multi-Region Latent Factor Analysis via Dynamical Systems (MR-LFADS), a sequential variational autoencoder designed to disentangle inter-regional communication, inputs from unobserved regions, and local neural population dynamics. We show that MR-LFADS outperforms existing approaches at identifying communication across dozens of simulations of task-trained multi-region networks. When applied to large-scale electrophysiology, MR-LFADS predicts brain-wide effects of circuit perturbations that were held out during model fitting. These validations on synthetic and real neural data position MR-LFADS as a promising tool for discovering principles of brain-wide information processing.
\pageonecrunch
\end{abstract}

\pageonecrunch
\section{Introduction} 
\label{sec:intro}

Large-scale neural recording technologies, such as high-density electrophysiology \cite{steinmetz2019distributed, siegle2021survey, international2023brain, chen2024brain, bennett2024shield} and calcium imaging \cite{sofroniew2016large, song2017volumetric, allen2017global}, now enable simultaneous recording of neural population activity across many brain regions. These advances have revealed that many sensory, cognitive, and motor processes engage spatially distributed networks in the brain  \citep{makino2017transformation, gilad2018behavioral, stringer2019spontaneous, musall2019single, allen2019thirst, jia2022multi}. Consequently, there has been growing interest in the design of data-driven {\it communication models} that seek to infer the pathways and content of communication between the recorded regions.

Accurately identifying inter-regional communication is challenging for at least {\it four} reasons \cite{biswas2020theoretical, kang2020approaches, keeley2020modeling, perich2020rethinking, semedo2020statistical, kass2023identification}. {\it First}, communication signals are not directly observed in multi-region recordings. Although some recorded neurons might project to other recorded regions, the identity and targets of these projection neurons are typically unknown. {\it Second}, models may need to account for inputs to the recorded brain regions from other regions that were not recorded during the experiment. {\it Third}, models should faithfully reconstruct activity within each recorded region by accounting for communication between recorded regions, inputs from unrecorded regions, and local neural population dynamics---capturing complex features such as structured trial-to-trial variability \cite{goris2014partitioning} and nonlinear, nonstationary, and state-dependent population dynamics  \cite{shenoy2013cortical, vyas2020computation, duncker2021dynamics, durstewitz2023reconstructing}. {\it Fourth}, accurate reconstruction of the recorded data does not guarantee accurate inference of the underlying communication. Many different models may sufficiently explain the recorded data, leading to ambiguities about which, if any, should be trusted for scientific interpretation.

In this work, we introduce Multi-Region Latent Factor Analysis via Dynamical Systems \definep{\mrlfads}, a multi-region communication model that directly addresses all of the challenges outlined above. \mrlfads is a probabilistic model that represents each recorded region with a distinct set of stacked recurrent neural networks \definep{RNNs} that capture the region's potentially nonlinear and nonstationary population dynamics. \mrlfads represents communication between observed regions and inputs from unobserved regions as disentangled sets of latent variables. Structured information bottlenecks encourage the model to infer inputs from unobserved regions only when their effects cannot be explained by communication among the recorded regions. \mrlfads infers single-trial initial conditions and time-varying inputs that together account for trial-to-trial variability in the recorded activity. This automatic inference of inputs eliminates the need to manually specify input signals, thereby avoiding strong, difficult-to-validate assumptions about how external signals influence each region. Finally, \mrlfads constrains communication to originate from model-reconstructed neural activity, rather than from more flexible latent representations---a design choice that, as we will show, enables more accurate inference of communication without sacrificing the quality of data reconstruction.

To evaluate MR-LFADS, we developed 37 synthetic multi-region datasets that capture real-world challenges in communication modeling across a range of neuroscience-relevant scenarios. On these datasets, \mrlfads consistently outperforms existing models in recovering the pathways and content of communication. Through targeted ablations of key design features, we demonstrate that these features indeed improve the identification of communication. We then applied \mrlfads to multi-region electrophysiological recordings in mice performing a decision-making task \cite{chen2024brain}. In a subset of trials that were held out during model fitting, photoinhibition was applied to the anterior lateral motor cortex. \mrlfads predicted the brain-wide effects of these circuit perturbations, suggesting that \mrlfads inferred an accurate account of inter-regional communication. Moreover, \mrlfads infers consistent communication across multiple training runs from different random initializations, demonstrating its robustness and reliability in real-data settings. 

\section{Related Work}
\label{sec:related-work}

Existing communication models can be broadly categorized as either {\it static} or {\it dynamic}. {\it Static methods} predict each timestep of neural activity in a target region from a corresponding timestep of activity in one or more source regions and then interpret predictive source activity as inter-regional communication \cite{kaufman2014cortical, perich2018neural, ruff2019simultaneous, veuthey2020single}. Reduced-rank regression \definep{RRR} is a prominent static technique that models target-region activity as a low-rank linear function of the source-region activity \cite{semedo2019cortical, macdowell2025multiplexed}. While static methods are straightforward to fit and interpret, they are typically limited to capturing instantaneous, linear dependencies. Consequently, they do not readily account for nonlinear or nonstationary relationships, or temporal structure that may arise due to neural population dynamics \cite{vyas2020computation}.

{\it Dynamic methods} explicitly model temporal dependencies using, for example, switching linear dynamical systems \definep{SLDS} \cite{linderman2016recurrent}, RNNs \cite{perich2020inferring}, or Gaussian processes \cite{yu2008gaussian, gokcen2022disentangling, gokcen2024uncovering}. We will pay particular attention to two such techniques that, like \mrlfads, are coupled nonlinear dynamical systems, with each dynamical system representing one recorded region: multi-population sticky recurrent SLDS \definep{\mpsrslds} \cite{glaser2020recurrent}, and Multi-Region Switching Dynamical Systems \definep{\mrsds} \cite{karniol2024modeling}. In \mpsrslds each region is modeled by an SLDS, while in \mrsds each is modeled as a switching nonlinear dynamical system. 

While these existing approaches to communication modeling address some of the challenges outlined in  \Cref{sec:intro}, none, to our knowledge, address all four challenges. In particular, none of these existing methods support inferring inputs from unobserved brain regions. This functionality is crucial because inputs from unobserved regions might influence the target region's population dynamics and modes of communication. Some approaches attempt to account for unobserved inputs by explicitly providing task-related signals as inputs to each region or by removing condition averages and modeling the residual single-trial neural activity. However, these manual strategies impose strong assumptions about the content and targets of input signals. As we will show, misspecifying such inputs risks confounding inferred population dynamics and communication, leading to models that accurately reconstruct neural activity through incorrect mechanisms. 

To address such unobserved inputs, \citet{pandarinath2018inferring} introduced Latent Factor Analysis via Dynamical Systems \definep{LFADS}, a sequential variational autoencoder \definep{sVAE} for modeling single-trial neural population dynamics within a single recorded brain region. LFADS jointly identifies a nonlinear dynamical system, implemented as an RNN, along with the single-trial initial conditions and time-varying unobserved inputs needed to drive the system to reconstruct single-trial neural population recordings. We henceforth use \define{SR-LFADS} to refer to a single-region LFADS model. MR-LFADS builds on this foundation to support multi-region modeling with explicit disentangling of communication, unobserved inputs, and local population dynamics. 




\begin{figure*}[ht!]
\vskip -.05in
\begin{center}
\centerline{\includegraphics[width=1.9\columnwidth]{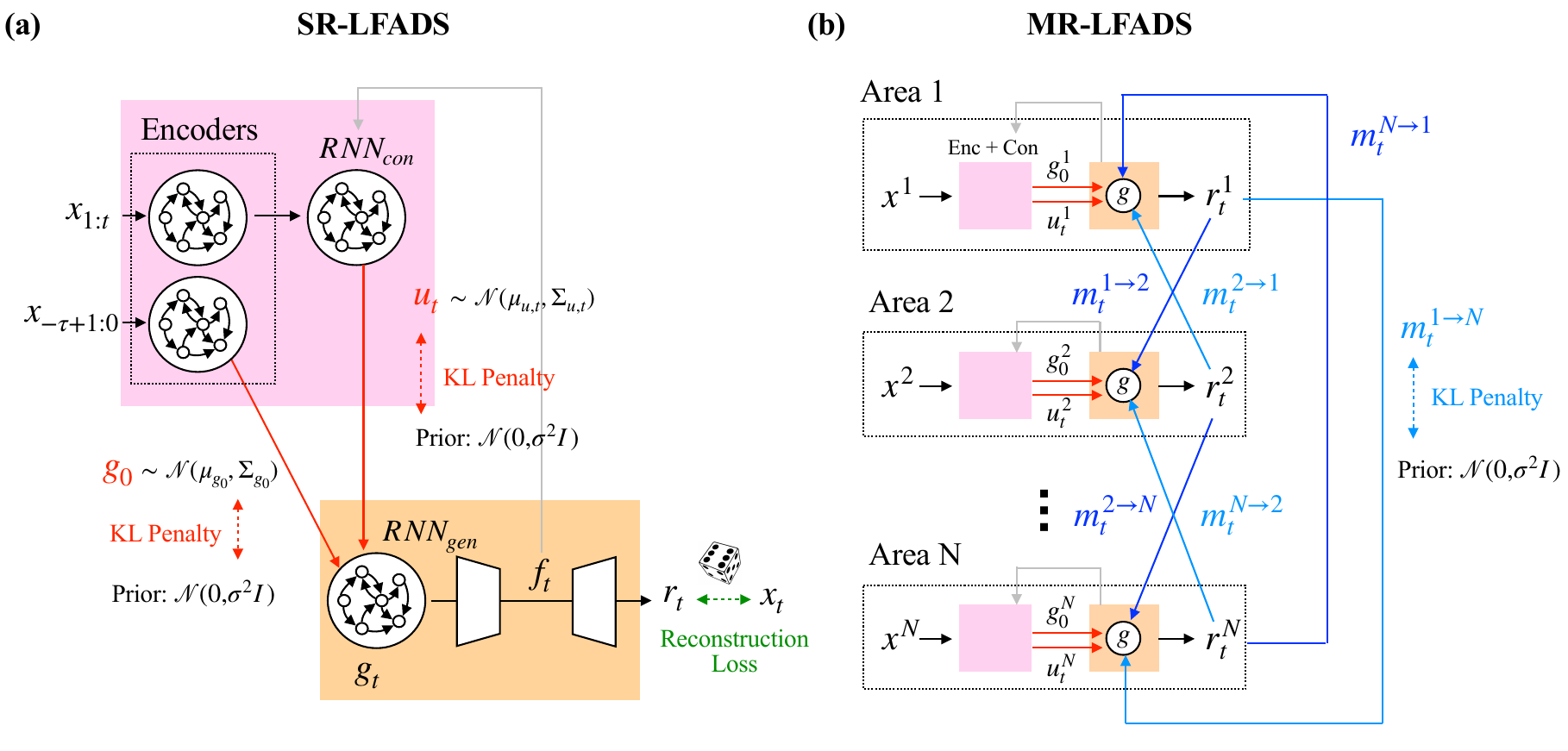}}
\vspace{-3.5ex}
\caption{\mrlfads architecture. (a) Single-region LFADS, as adapted for this work. (b) \mrlfads with $N$ regions. KL penalties from SR-LFADS in panel (a) are included in \mrlfads, but are omitted in the diagram for clarity.}
\label{fig:mrlfads}
\end{center}
\vspace{-5ex}
\doublefigline
\vspace{-4ex}
\end{figure*}

\section{Multi-Region LFADS (\mrlfads)}
\label{sec:mrlfads}

\mrlfads is composed of a set of SR-LFADS modules (\Cref{fig:mrlfads}a) that interact through constrained communication channels (\Cref{fig:mrlfads}b). At a high level, \mrlfads is a coupled set of driven nonlinear dynamical systems that are jointly trained to reconstruct all single trials of a multi-region dataset. Each recorded brain region $i$ is modeled as a  dynamical system that attempts to reconstruct the region-$i$ recorded neural activity \(\data{t}{i}\) at each time $t=1,\dots,T$. Each region-$i$ dynamical system evolves from a single-trial initial state \(\genstate{0}{i}\) and is driven by (1) single-trial time-varying communication messages \(\msg{t}{j}{i}\) from other recorded brain regions \(j\) and (2) single-trial time-varying inferred inputs $\infinput{t}{i}$, representing input from unobserved brain regions.

\textbf{Notation.} All time-indexed variables and parameters are also indexed by trial, though we omit trial indices for notational simplicity. We use \(t_1\):\(t_2\) to denote the inclusive sequence of integers \(\{t_1, t_1+1, \dots , t_2\}\).  We use \(W^i(x)\hspace{-.25em}:=\hspace{-.2em}W^ix+b^i\) to denote an affine transformation with weights $W^i$ and offsets $b^i$.

\textbf{Generative Model.} \mrlfads treats all initial states $\genstate{0}{i}$, communications $\msg{t}{j}{i}$, and inferred inputs $\infinput{t}{i}$ as latent variables. The prior distributions over these latent variables are modeled as:
\begin{equation}
\genstate{0}{i}, \infinput{t}{i}, \msg{t}{j}{i}  \sim \mathcal{N}(0, \sigma^2I) \label{eq:prior}
\end{equation} 
Given these quantities, the neural population dynamics in region $i$ are modeled by a ``generator" gated recurrent unit \definep{GRU} network, $\genrnn{i}$, with internal states $\genstate{t}{i}$ that evolve according to:
\begin{equation}\label{eq:gen}
\begin{aligned}
    & \genstate{t}{i} = \genrnn{i}\big(\genstate{t-1}{i}, \big[\{\msg{t}{j}{i}\}_{j \neq i}; \infinput{t}{i} \big] \big)
\end{aligned}
\end{equation} 
A set of region-$i$ factors $\factors{t}{i}$ are defined as an affine readout from the corresponding generator states:
\begin{equation}\label{eq:fac}
\begin{aligned}
    \factors{t}{i} = W_{\factors{}{}}^i(\genstate{t}{i})
\end{aligned}
\end{equation}
These factors are then transformed into parameters of time-varying output distributions in a manner dependent on the nature of the neural recordings. For continuous-valued observations, as in calcium imaging, a Gaussian or zero-inflated Gamma distribution may be appropriate \cite{zhu2022deep}. In the synthetic-data experiments of \Cref{sec:synth}, we apply a Gaussian output distribution:
\begin{equation} \label{eq:gaussian}
P(\data{t}{i} \mid  \genstate{0}{i}, \infinput{1:t}{i}, \{\msg{1:t}{j}{i}\}_{j \neq i}) = \mathcal{N}( \rates{t}{i}, \Sigma_{\rates{}{},t}^{i}) 
\vspace{-3pt}
\end{equation}
\begin{equation} \label{eq:gaussianparams}
\rates{t}{i} = {W}_{\rates{}{}}^{i}(\factors{t}{i}) \quad \quad
\Sigma_{\rates{}{},t}^{i} = \mbox{diag} \left(\exp(W_{\sigma_{\rates{}{}}}^{i}(\factors{t}{i})) \right)
\end{equation} 
where $\rates{t}{i}$ and $\Sigma_{\rates{}{},t}^i$ are the region-$i$ predicted mean and covariance of the time-$t$ recorded neural activity, respectively, and are each computed via separate affine transformations,  $W_{\rates{}{}}^i$ and $W_{\sigma_{\rates{}{}}}^i$. For spike count observations, as modeled in the electrophysiology experiments of \Cref{sec:ephys}, we apply a Poisson output distribution:
\begin{equation} \label{eq:poiss}
    P(\data{t}{i} \mid \cdot) = \mbox{Poisson}(\rates{t}{i}) \quad \quad
    \rates{t}{i} = \exp \big(W_{\rates{}{}}^{i}(\factors{t}{i}) \big)
\end{equation} 
where the exponential nonlinearity ensures non-negative predicted firing rates $\rates{t}{i}$.

\vspace{1pt}
\textbf{Inference Model.} Following VAE conventions \cite{kingma2013auto}, \mrlfads approximates the intractable true posterior distributions over the latent variables using variational posteriors, denoted $q(\cdot | \cdot)$. 

\mrlfads defines the approximate posteriors over communication messages from observed regions $j$ to $i$ as Gaussian distributions:
\begin{equation} \label{eq:msg}
q(\msg{t}{j}{i} \mid \data{1:t}{j}) = \  q(\msg{t}{j}{i} \mid \rates{t}{j}) = \mathcal{N}(\mu^{j\rightarrow i}_{m,t}, \Sigma^{j\rightarrow i}_{m,t})
\end{equation} 
 with parameters derived from the region-$j$ predicted firing rates $\rates{t}{j}$:
\begin{equation} \label{eq:msgparams}
\hspace{-10pt}
\begin{aligned}
\mu^{j\rightarrow i}_{m,t} &= {W}^{j\rightarrow i}_{\mu_m}(\rates{t}{j}) \\
\Sigma^{j\rightarrow i}_{m,t} &= \mbox{diag} \Big(
\exp\big(W_{\sigma_m}^{j\rightarrow i}(\rates{t}{j})\big)
\Big)
\end{aligned}
\end{equation}
Constraining communication to be derived from $\rates{t}{j}$ anchors it to the neural recordings and in doing so reduces ambiguity in system identification. We refer to this rate-based communication model as \define{\mrlfadsr}. In \Cref{sec:synth}, we also explore generator-based \define{\mrlfadsg} and factor-based \define{\mrlfadsf} communication models, which replace all instances of $\rates{t}{j}$ in \Cref{eq:msgparams} with $\genstate{t}{j}$ and $\factors{t}{j}$, respectively. 

\pagebreak

Approximate posteriors over the region-$i$ generator initial states $\genstate{0}{i}$ and inferred inputs (from unobserved brain regions) $\infinput{t}{i}$ are defined as the following Gaussian distributions:
\begin{align}
    q(\genstate{0}{i} \mid \data{-\tau:0}{i}) &= \mathcal{N} (\mu^i_{\genstate{0}{}}, \Sigma^i_{\genstate{0}{}}) \label{eq:qg} \\ 
    q(\infinput{t}{i} \mid \data{1:t}{i}) &= \mathcal{N} (\mu^i_{\infinput{}{},t}, \Sigma^i_{\infinput{}{},t}) \label{eq:qu}
\end{align}
where the mean and covariance parameters are computed from the corresponding conditioning recorded neural activity $\data{}{i}$ via a set of region-$i$-specific ``encoder'' and ``controller'' GRU networks (see \hyperref[app:mrlfads]{Appendix A.1}). Our approach here slightly modifies the original SR-LFADS specification, which allows acausal inference via a bidirectional encoder network that processes each entire $T$-timestep neural recording $\data{1:T}{}$ to infer $\genstate{0}{}$ and each element of $\infinput{1:T}{}$. In contrast, we infer $\genstate{0}{i}$ (\Cref{eq:qg}) using a bidirectional encoder applied only to past neural activity \( \data{-\tau:0}{i} \), preserving causality, and we infer \( \infinput{t}{i} \) (\Cref{eq:qu}) using a unidirectional encoder RNN that processes \( \data{1:t}{i} \) in a strictly forward, causal manner. This formulation ensures that all predicted firing rates $\rates{t}{i}$, and thus all derived communication signals \( \msg{t}{j}{i} \), are inferred causally from neural activity recorded up to time \( t \). 

\textbf{Model Fitting.} Following VAE conventions, \mrlfads is trained by maximizing the evidence lower bound \definep{ELBO}, a variational lower bound on the data log-likelihood. The ELBO is a sum of two terms: (1) the expected log-likelihood
\begin{equation} \label{eq:reconstruction}
\sum_{t=1}^T\mathbb{E}_q \big[ \log P(\{\data{t}{i}\} \mid \{\genstate{0}{i}\},\{\infinput{t}{i}\}, \{\msg{t}{j}{i} \}) \big]
\end{equation}
and (2) the negative Kullback-Leibler \definep{KL} divergence $D_\text{KL}$ between the approximate posteriors (\Crefrange{eq:msg}{eq:qu}) and the priors (\Cref{eq:prior}) over the latent variables. 

The expected log-likelihood measures reconstruction accuracy and is estimated by running samples from the approximate posteriors (\Crefrange{eq:msg}{eq:qu}) through the generative model to evaluate the experiment-dependent output distributions from \Crefrange{eq:gaussian}{eq:poiss}. The $D_\text{KL}$ term acts as a regularizing information bottleneck on the latent variables. To control this regularization, we allow rescaling of the $D_\text{KL}$ term \cite{higgins2017beta, keshtkaran2022large}. Noting that the $D_\text{KL}$ term decomposes into contributions from the three sets of \mrlfads latent variables $\{\genstate{0}{i}\}$, $\{\infinput{t}{i}\}$, and  $\{\msg{t}{j}{i}\}$, we weight each contribution differently and treat the weights $(\klweight{\genstate{0}{}}, \klweight{u}, \klweight{m})$ as hyperparameters. To encourage \mrlfads to infer inputs from unobserved regions only when that information cannot be obtained as communication from an observed region, we propose a structured KL bottleneck with $\klweight{u} = 10 \klweight{m}$. Other choices of KL regularization structure might be appropriate if \textit{a priori} knowledge is available about information flow or anatomical connectivity between recorded regions. See \hyperref[app:mrlfads]{Appendix A.1} for further detail on MR-LFADS.

\section{Results I: Synthetic Multi-Region Datasets}
\label{sec:synth}

Here, we evaluate \mrlfads' ability to recover the ground truth inter-regional communication across a broad range of synthetic multi-region datasets. Each dataset was generated by a unique data-generating network \definep{DGN}: an ensemble of noisy RNN modules jointly trained to perform a specified cognitive neuroscience task, with each module representing a distinct brain region. Prior to training, we explicitly specified the presence or absence of directed, low-rank communication channels between each region pair. 

In Experiments 1 and 2, we manually designed the DGNs to impose specific challenges outlined in \Cref{sec:intro}. In Experiment 3, we randomly generated dozens of DGNs, each trained to perform a randomly selected cognitive neuroscience task. Across all experiments, we treated each module's hidden-unit activity as recorded neural activity and retained all external inputs, inter-module connectivity, and communication signals as ground truth. These ground truth quantities are key to evaluating communication models but are typically not directly observable in real-data settings. 


To benchmark \mrlfads, we compare it to three established communication modeling techniques: RRR (see \hyperref[app:rrr]{Appendix A.2}), \mpsrslds, and \mrsds (see \Cref{sec:related-work}). To demonstrate the importance of specific \mrlfads design features, we also compare against ablated \mrlfads variants that selectively exclude those features. We focus evaluation on each method's ability to recover a causal model of each DGN---including both the {\it pathways} and {\it content} of inter-regional communication---given a multi-region dataset generated by the DGN. 

To assess communication {\it pathways}, we consider recovery of an ``effectome''  \cite{pospisil2024fly} describing the causal flow of effects along the inter-regional connectome. We represent this effectome as a matrix with each element $(i,j)_{i \neq j}$ indicating the volume of directed communication flow from region $j$ to $i$. The effectome reflects both the inter-regional connectivity and the magnitude of communication flow over each directed connection. For each dataset, we compare model-inferred effectomes to the ground truth effectome by computing cosine similarity $\scos$ between the vectorized effectome matrices.

To assess communication {\it content}, we quantify how well the model-inferred messages capture the information in the ground truth messages. Specifically, we apply linear regression to predict the ground truth messages $\msg{t}{j}{i}$ from the inferred messages $\mu_{m,t}^{j \rightarrow i}$ (\Cref{eq:msg}), and we report a cross-validated coefficient of determination, denoted $R^2(\mu_m^{j \rightarrow i}, \msg{}{j}{i} )$. For MR-LFADS, we use an analogous procedure to compare model-inferred inputs $\mu_{u,t}^i$ (\Cref{eq:qu}) to ground truth external inputs. This comparison is not applicable to RRR, \mpsrslds, or \mrsds, as these methods do not infer inputs from unobserved regions. Additional details on experiments, model hyperparameters, and evaluation metrics are provided in \hyperref[app:memory]{Appendix B-D}.

\subsection{Experiment 1: Inferring Unobserved Inputs} 
\begin{figure}[hb!]
\vspace{-5ex}
\figline
\includegraphics[width=\columnwidth]{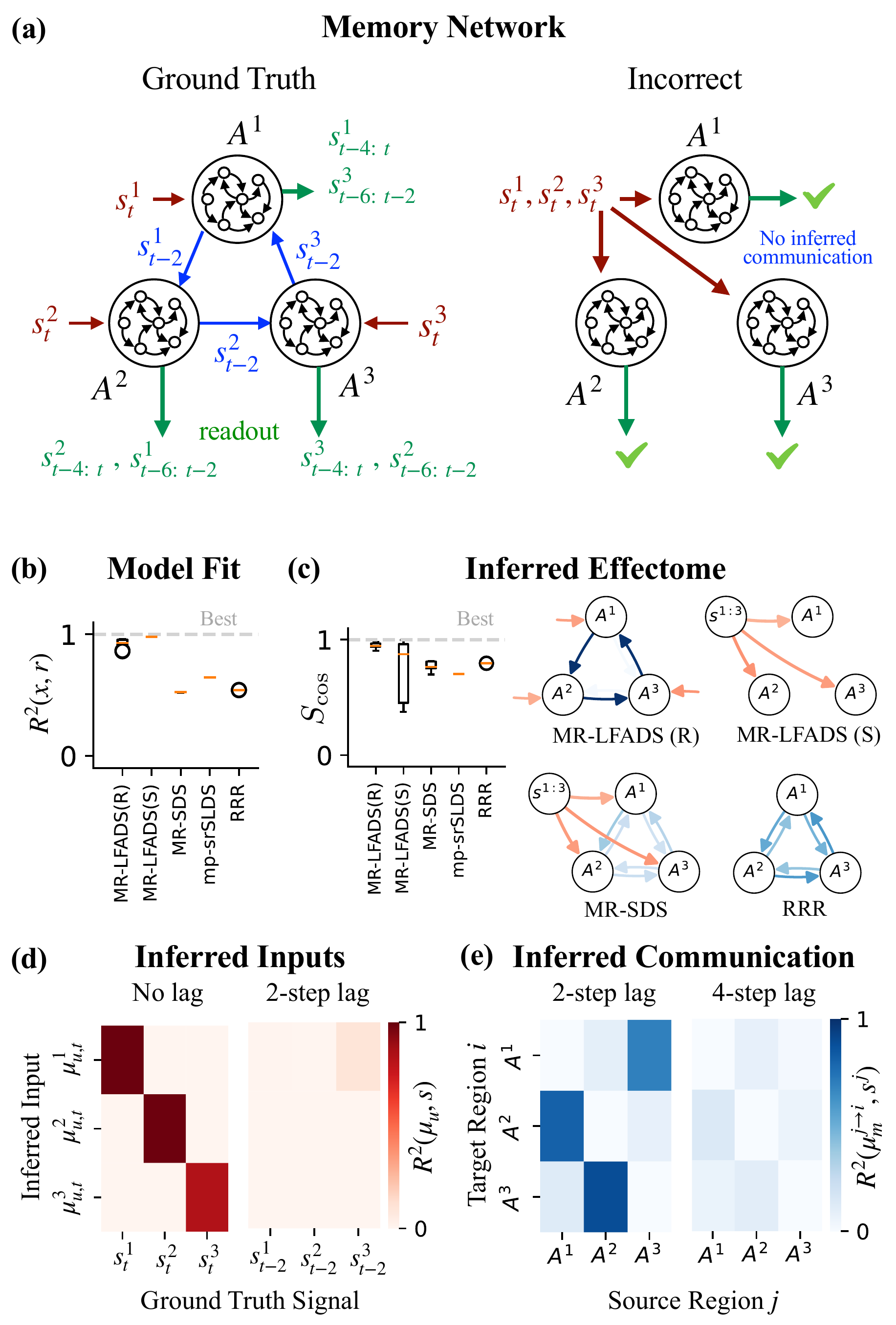}
\vspace{-5ex}
\caption{Experiment 1. (a) \textit{Left:} DGN setup. Each region, area $A^i$, receives a private stimulus $\stim{t}{i}$ (red) and communicates a two-step delayed version $\stim{t-2}{i}$ (blue) to a downstream region. Each region is trained to recall (green) the last five time steps of its private stimulus and its received communication. \textit{Right:} Potential incorrect communication model that accurately reconstructs these synthetic neural data. (b) $R^2$ scores for data reconstruction. Box plots show distributions across $10$ models fit from distinct random initializations (seeds). Boxes represent the interquartile range (IQR), and whiskers extend to the most extreme points within $1.5$ IQR from the quartiles. (c) \textit{Left:} Cosine similarity between inferred effectomes and ground truth. \textit{Right:} Example fitted models. Color intensity indicates the relative message norm, computed by concatenating all multidimensional messages across trials and time, taking the 2-norm, then normalizing across communication channels. $s^{1:3}$ indicates the ground truth input. (d) $R^2$ of linear prediction of ground truth stimulus inputs (with time lag $\in \{0, 2\}$) from inferred inputs. (e) $R^2$ of linear prediction of ground truth messages (with time lag $\in \{2, 4\}$) from inferred messages.}
\label{fig:exp1}
\end{figure}

This experiment demonstrates that inferring inputs, rather than manually specifying inputs, enables more accurate identification of inter-regional communication. To evaluate design implications related to input specification, we designed a DGN that implements a dynamical memory function. Each region of this ``memory network'' receives unique stimulus information from one observed region and one unobserved region, and is tasked with remembering a recent history of those signals (\Cref{fig:exp1}a, left). With each region of the DGN receiving information from both observed and unobserved sources, this setup poses the challenge of disentangling whether each signal arises due to communication or due to external input. A common modeling choice is to provide all known external inputs to all model regions and to let model fitting determine which inputs are needed by each region. However, in this case, such manual input specification can result in a model completely forgoing communication (\Cref{fig:exp1}a, right) because the manually specified inputs contain the information that was actually transmitted as communication in the DGN.

We evaluate \mrlfadsr, which does not use the stimulus signals during training or evaluation but rather automatically infers external inputs in an unsupervised manner. We also evaluate an ablated \mrlfads variant that does not automatically infer inputs. Termed \mrlfadss, this model receives all stimulus signals as manually specified inputs to each region's generator $\genrnn{i}$ (replacing $\infinput{t}{i}$ with $\stim{t}{1:3}$ in \Cref{eq:gen}). Likewise, \mrsds and \mpsrslds receive all stimulus signals as manually specified inputs, whereas RRR neither receives nor infers external inputs.

The \mrlfads variants (R and S) reconstructed the simulated multi-region activity more accurately than \mrsds, mp-rSLDS, and RRR (\Cref{fig:exp1}b, \Cref{fig:exp1s}a). Critically, \mrlfadsr accurately infers the effectome (\Cref{fig:exp1}c, left, \Cref{fig:exp1s}b). By contrast, all models that are manually provided stimulus signals (\mrlfadss, \mrsds, mp-rSLDS) infer less accurate effectomes and demonstrate the failure mode mentioned above, forgoing communication and instead relying on the specified inputs to provide the corresponding signals. RRR also infers a less accurate effectome. These results suggest that manually specifying inputs can discourage models from utilizing---and thus identifying---communication. By automatically inferring inputs, \mrlfadsr avoids this failure mode. 

Next, we assess the accuracy of inferred inputs and messages. In the ground truth DGN at time $t$, region-$i$ receives only $\stim{t}{i}$ and $\msg{t}{j}{i}$. An accurate communication model should therefore infer inputs and messages that encode only these time-$t$ quantities. However, due to structure of the memory task, the DGN's region-$i$ time-$t$ activity contains information about $\stim{t-4:t}{i}$ and $\msg{t-4:t}{j}{i}=\stim{t-6:t-2}{j}$. To reconstruct the data, a communication model must account for how this time-lagged information becomes represented in region-$i$ at time-$t$. \mrlfadsr correctly infers the current timestep ground truth stimuli $\stim{t}{i}$ as inputs to each region (\Cref{fig:exp1}d, left), while correctly avoiding inferring time-lagged versions of those stimuli (\Cref{fig:exp1}d, right), despite their utility for data reconstruction. Similarly, \mrlfadsr accurately recovers the current timestep messages $\msg{t}{j}{i}$ (\Cref{fig:exp1}e, left) and avoids incorrectly inferring time-lagged versions (\Cref{fig:exp1}e, right). By contrast, all other models infer no communication or communication with incorrect temporal lags (\Cref{fig:exp1s}c). See \hyperref[app:msg]{Appendix B.2} for further details.

Taken together, these results indicate that \mrlfadsr learned region-specific population dynamics consistent with those that implement the memory functions in the ground truth DGN. Moreover, these results demonstrate the unique ability of \mrlfadsr to disentangle region-specific external inputs, communication between recorded regions, and local population dynamics, all in an unsupervised manner that mitigates biases associated with manual specification of external inputs. See \hyperref[app:kl_sparsity]{Appendix E.1} for further analyses linking this disentangling \cite{miller2024cognitive} to \mrlfadsr's structured information bottlenecks.

\subsection{Experiment 2: Data-Constrained Communication} This experiment demonstrates the implications of message-inference design choices. We designed \mrlfadsr to infer messages as affine functions of the source-region predicted firing rates $\rates{t}{i}$ (\Crefrange{eq:msg}{eq:msgparams}), which are tied to the observed source-region neural activity $\data{t}{i}$ through the data reconstruction term in the ELBO (\Cref{eq:reconstruction}). This data-constrained communication architecture contrasts with that of \mrsds and \mpsrslds, which infer communication as a function of less-constrained, source-region latent dynamical states (see \Cref{table:design}). To directly evaluate the implications of this design choice, all within the \mrlfads framework, we designed model variants with less-constrained factor-based and generator-based communication, termed \mrlfadsf and \mrlfadsg, respectively.

\begin{figure}[hb!]
\vspace{-4ex}
\figline
\includegraphics[width=\columnwidth]{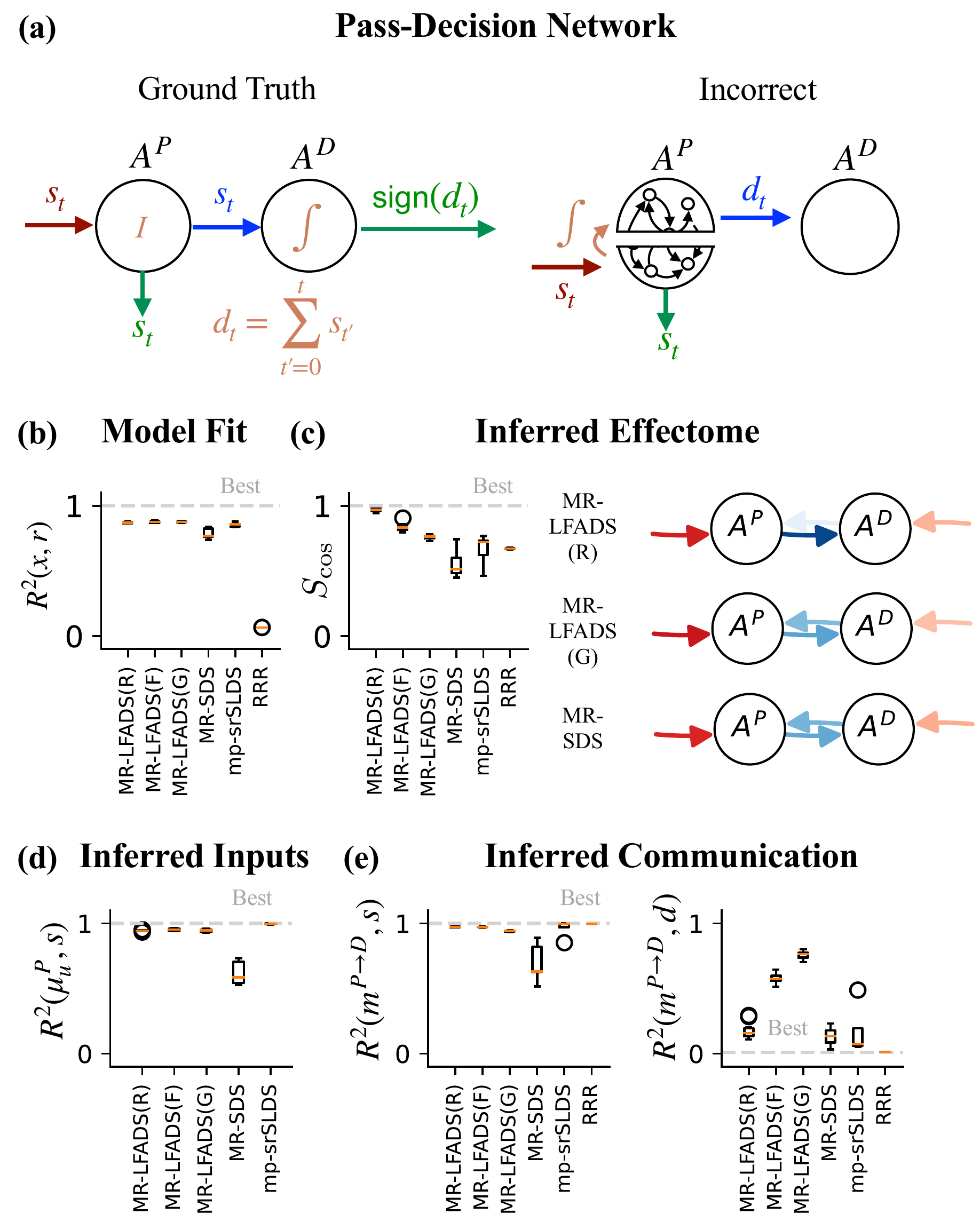}
\vspace{-5ex}
\caption{Experiment 2. (a) \textit{Left:} DGN setup, with stimulus (red), communication (blue), trained readouts (green), and computations (\textit{I}: identity; $\int$: integration). \textit{Right:} Potential failure mode for a learned model. (b) $R^2$ scores for data reconstruction. (c) \textit{Left:} Cosine similarity between inferred effectomes and ground truth. \textit{Right:} Example fitted models. Color intensity indicates relative message norm. (d) $R^2$ of linear prediction of ground truth input $s$ to region $P$, from inferred inputs. (e) \textit{Left:} $R^2$ when predicting ground truth messages $\msg{}{P}{D}$$=$$\stim{}{}$ from inferred messages, $\mu_m^{P \rightarrow D}$. \textit{Right:} $R^2$ when predicting the decision variable $\decision{}$ from $\mu_m^{P \rightarrow D}$, indicating mislocalization of integration.}
\label{fig:exp2}
\end{figure}

To highlight the significance of this design choice, we evaluate models on data from a two-region ``pass-decision'' DGN that computes perceptual decisions based on time-varying sensory evidence (\Cref{fig:exp2}a, left). An upstream region, area $A^P$, receives a white noise stimulus $\stim{t}{}$ and is trained to {\it pass} that stimulus through to a readout, effectively learning an identity function routing input to output. A downstream {\it decision} region, area $A^D$, receives this routed stimulus as communication $\msg{t}{P}{D}$, integrates that stimulus over time into a decision variable $\decision{t}$, and reports $^\pm1$ choices indicating the sign of that decision variable \cite{mante2013context}. 

This setup again challenges models to disentangle external inputs, communication, and local dynamics---and in particular, accurately identifying and localizing the ground truth pass-through and integration computations. Though the integration dynamics are localized to $A^D$ in the DGN, an overly flexible model might instead learn integration in $A^P$\hspace{-.5em}, yielding accurate data reconstruction but mislocalizing the computation, e.g., if $A^P$ integrates $\stim{t}{}$ into $\decision{t}{}$ and communicates $\decision{t}{}$ as $m^{P\rightarrow D}$ to $A^D$ (\Cref{fig:exp2}a, right).

On this pass-decision dataset, all \mrlfads models achieved comparably high-fidelity data reconstruction, slightly outperforming \mrsds and \mpsrslds (\Cref{fig:exp2}b, \Cref{fig:exp2s}a). RRR reconstructed the data poorly, likely due to its inability to capture the timescale of the integration computation, i.e., $\decision{t}{}$ cannot be predicted from any single-timestep value $\stim{t'}{}$. 

\mrlfadsr inferred the most accurate effectome (\Cref{fig:exp2}c, \Cref{fig:exp2s}b), identifying $A^P$$\rightarrow$$A^D$ communication but not $A^D$$\rightarrow$$A^P$. By contrast, all other models, including \mrlfadsf and \mrlfadsg, identified spurious $A^D$$\rightarrow$$A^P$ communication. 

\pagebreak

All \mrlfads variants correctly inferred inputs to $A^P$ that encoded $\stim{t}{}$ (\Cref{fig:exp2}d). Because \mrsds and \mpsrslds do not infer unsupervised inputs, we estimated their effective inputs by passing the manually specified inputs through their corresponding trained input mappings. In \mrsds, these effective inputs to $A^P$ carried markedly less information about $\stim{t}{}$ relative to \mrlfads and \mpsrslds. 

Accurate identification of communication requires inferred messages $\msg{t}{P}{D}$ to encode the stimulus $\stim{t}{}$. By contrast, $\msg{t}{P}{D}$ instead encoding the decision variable $\decision{t}$ would imply mislocalization of the integration dynamic (\Cref{fig:exp2}a, right). Only \mrlfadsr, \mpsrslds, and RRR correctly encoded $\stim{t}{}$ in $\msg{t}{P}{D}$ (\Cref{fig:exp2}e, left) without incorrectly encoding $\decision{t}$ (\Cref{fig:exp2}e, right). 

\begin{figure*}[hb!]
\vspace{-3ex}
\doublefigline
\vspace{.5ex}
\includegraphics[width=2.09\columnwidth]{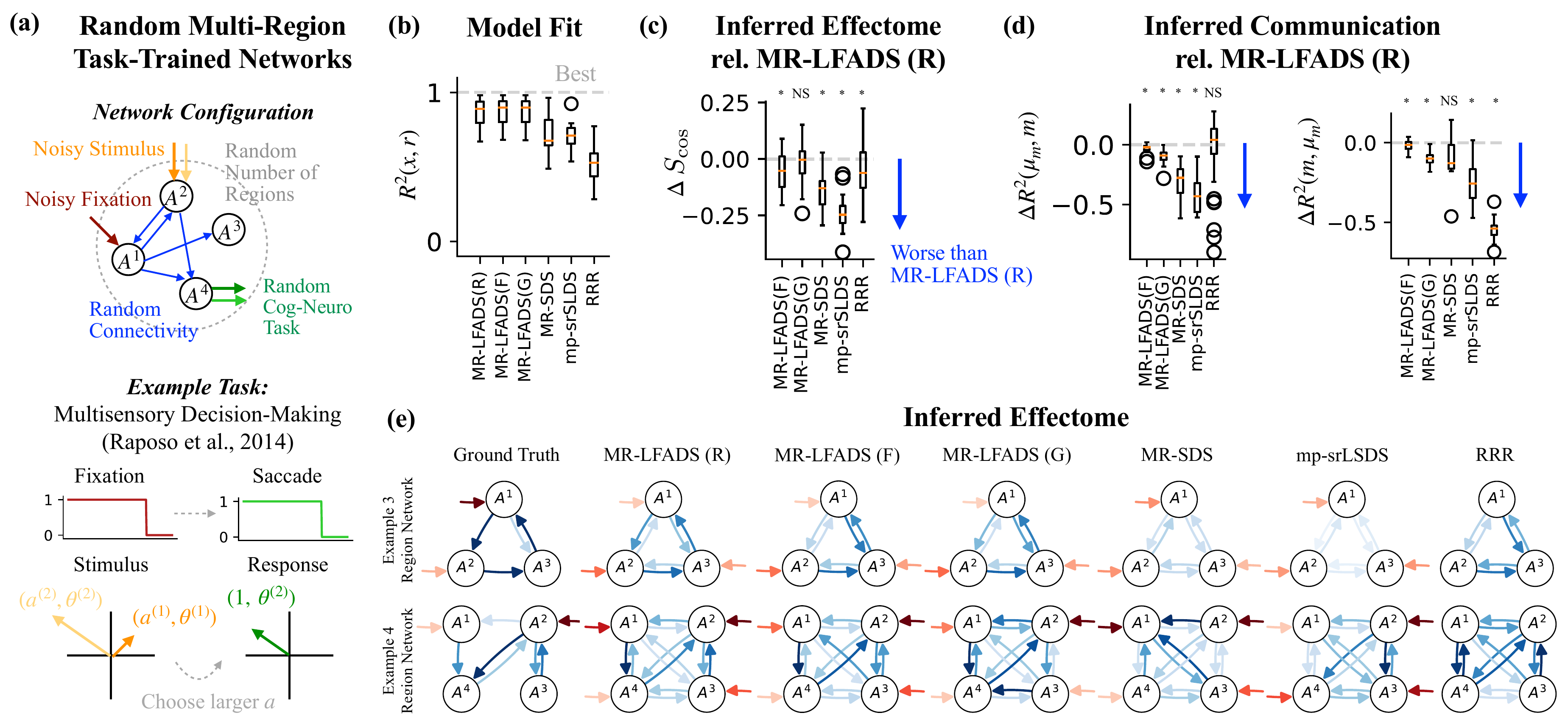}
\vspace{-5ex}
\caption{Experiment 3. (a) \textit{Top:} Each DGN is configured with a random number of areas ($3$ or $4$), random inter-regional connectivity, and is trained on a randomly selected task. \textit{Bottom:} Example task setup. On each trial, the DGNs received a noisy fixation stimulus $\stim{\text{fix}, t}{}$ and two noisy task stimuli, represented in polar coordinates as $\stim{t}{k} = (a_t^{(k)}, \theta_t^{(k)})$ for $k \in \{1, 2\}$.  DGNs were trained to process these inputs into task-dependent outputs: a response angle $\theta^\text{resp}$ and a task-dependent saccadic eye movement. Here, multisensory decision-making is depicted as an example task. When the fixation cue disappears, the output area $A^N$ must saccade and report the $\theta^{(i)}$ value corresponding to the stimulus with larger $a^{(i)}$. (b) $R^2$ scores for data reconstruction. Box plots describe distributions of values across $35$ DGN fits. (c) Cosine similarity ($\scos$) of inferred effectomes relative to ground truth, compared to that of \mrlfadsr, with $\Delta S_\text{cos} = S_\text{cos}^\text{model} - S_\text{cos}^\text{\mrlfadsr}$. One-tailed t-test p-values: $p = 0.00016$, $0.12$, $0.006$, $0.0$, $0.01$. NS: not significant. (d) \textit{Left:} $R^2$ scores, relative to \mrlfadsr, for linear prediction of ground truth messages from inferred messages. One-tailed t-test p-values: $p = 0.0$, $0.0$, $0.001$, $0.0$, $0.14$. \textit{Right:} $R^2$ scores, relative to \mrlfadsr, for linear prediction of inferred messages from ground truth messages. One-tailed t-test p-values: $p = 0.0004$, $0.0$, $0.08$, $0.0$, $0.0$. (e) Inferred effectomes from models fit to datasets from a three-region DGN (\textit{top}) and a four-region DGN (\textit{bottom}). We chose the datasets on which \mrlfadsr achieved its median $\scos$ scores (across the 35 generated datasets). Color intensity indicates relative message norm.}
\label{fig:exp3}
\end{figure*}

Taken together, these Experiment 2 results again demonstrate \mrlfadsr outperforming existing methods at disentangling external inputs, communication, and local population dynamics, without sacrificing data reconstruction. The comparisons to \mrlfadsf and \mrlfadsg specifically highlight the importance of \mrlfadsr's data-constrained communication, which improves identification of communication by reducing ambiguity inherent to overly flexible models. See \hyperref[app:sing_val]{Appendix E.2} for further analyses into this excessive flexibility.

\vspace{-1pt} 
\subsection{Experiment 3: Generalization Across Random Multi-Region Networks}
\vspace{-1pt} 

The previous experiments utilized datasets synthesized by DGNs designed to highlight specific failure modes of communication models. However, real brain-wide networks exhibit a broad range of architectures and computations. To assess generalization across such a broad range of settings, here we evaluate communication models on datasets generated by a wide variety of randomly configured multi-region DGNs (\Cref{fig:exp3}a, top), each trained to perform a randomly selected cognitive neuroscience task (\Cref{fig:exp3}a, bottom). We generated 35 multi-region datasets, each from a unique task-trained DGN consisting of three or four regions and randomized inter-regional connectivity. Tasks were drawn from the set described by \citet{yang2019task}, spanning multiple variants of decision-making, working memory, categorization, and inhibitory control (see \hyperref[app:yang]{Appendix D}). 

We fit \mrlfadsrfg, along with \mrsds, \mpsrslds, and RRR, to each of these datasets. Aggregating results across all datasets, the \mrlfads models achieved the best data reconstruction, which was indistinguishable across model variants (\Cref{fig:exp3}b, \Cref{fig:exp3s1}). \mrlfadsr and \mrlfadsg inferred the most accurate effectomes, with statistically indistinguishable $\scos$ distributions (\Cref{fig:exp3}c). To evaluate the accuracy of inferred message content, we attempted to linearly decode ground truth messages from inferred messages and quantified accuracy using the  $R^2$ scores of these predictions (\Cref{fig:exp3}d, left). We also performed the reverse, predicting inferred messages from ground truth, and interpreted lower $R^2$ scores as an indication that inferred messages contained additional information beyond that present in the ground truth (\Cref{fig:exp3}d, right). \mrlfadsr was the only model that performed best across both of these metrics. Inferred effectomes from two example datasets are shown in \Cref{fig:exp3}e. Taken together, these results demonstrate that \mrlfadsr outperforms existing communication models across a broad range of neuroscience-relevant synthetic multi-region datasets.

\section{Results II: Multi-Region Electrophysiology} \label{sec:ephys}

\begin{figure*}[hb!]
\vspace{-3ex}
\doublefigline
\vspace{1ex}
\includegraphics[width=2.08\columnwidth, trim=11pt 0 0 0 0,clip]{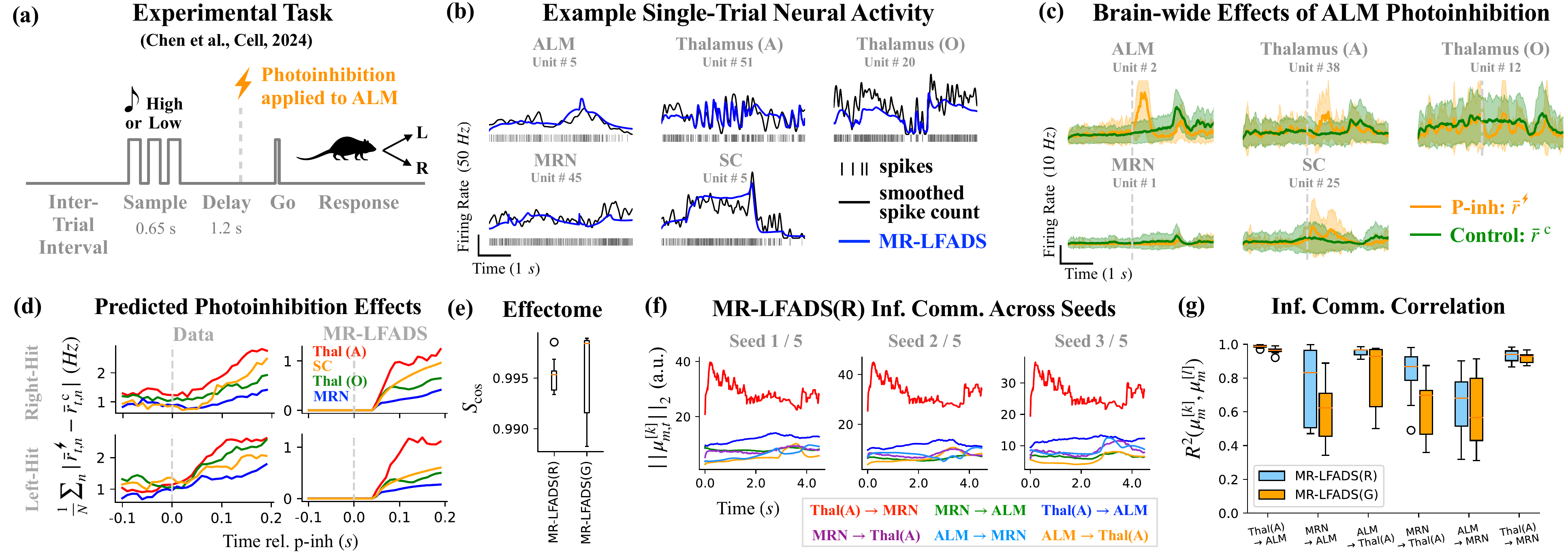}
\vspace{-5ex}
\caption{\mrlfadsr applied to multi-region, high-density electrophysiology.
(a) Mice receive an auditory stimulus (``sample'') and respond by licking left or right (``response'').  (b) \mrlfads single-trial predicted firing rates in held-out control trials (blue), recorded spike times (black vertical ticks), and smoothed, binned spike counts (black; causal exponential filter) for example neurons in each modeled brain region.
(c) Condition-averaged smoothed spike counts of example neurons in control and photoinhibition trials (left-hit condition). Shaded regions indicate standard deviation across trials.
(d) Photoinhibition-related changes in population activity, observed experimentally (left) and predicted by \mrlfads (right).
(e) Cosine similarity of inferred effectomes across models with different random initializations (seeds).
(f) Message norms inferred by \mrlfadsr for all connections across example seeds $k$.  
(g) Correlation of inferred message norms across pairs of seeds $(k,l)$. Box plots indicate results across $5$ models fit from distinct random initializations.}
\label{fig:ablation}
\end{figure*}

Here, we apply \mrlfadsr to large-scale electrophysiological data from multiple simultaneously recorded Neuropixel probes in mice performing a decision-making task (\Cref{fig:ablation}a) \cite{chen2024brain}. We test whether \mrlfads can predict the effects of causal circuit perturbations that were held out from training. We also assess the reliability of inferred inter-regional communication across random initializations, comparing \mrlfadsr to \mrlfadsg.

We trained a 5-region \mrlfadsr model on simultaneously recorded population activity from anterior lateral motor cortex \definep{ALM}, midbrain reticular nucleus \definep{MRN}, superior colliculus \definep{SC}, thalamic regions with strong, reciprocal connections to ALM \definep{Thal(A)}, and other thalamic regions \definep{Thal(O)}. In a subset of trials excluded from training, ALM was transiently photoinhibited (\Cref{fig:ablation}a). \mrlfads was trained only on unperturbed (``control'') trials, with photoinhibition trials held out for validation. See \hyperref[app:exp]{Appendix F} for further detail.

We first verified that \mrlfads accurately reconstructed multi-regional neural activity in held-out control trials (\Cref{fig:ablation}b). We then adapted \mrlfadsr to predict the effects of ALM photoinhibition on the remaining recorded regions (\Cref{fig:ablation}c). To mimic ALM photoinhibition \textit{in silico}, we ablated \mrlfads communication from ALM to the other regions by zeroing all outgoing messages from ALM. We summarized the temporal influence of ALM photoinhibition by computing the differences between condition-averaged population activity in photoinhibition ($\bar{r}_t^{\lightning}$) and control ($\bar{r}_t^{c}$) trials. \mrlfads predicted these photoinhibition effects (\Cref{fig:ablation}d), despite never seeing photoinhibition trials during training. Namely, \mrlfads predicted that Thal(A) would be most affected by ALM photoinhibition, MRN least affected, and SC and Thal(O) intermediately affected. These results demonstrate that \mrlfads learned a model of inter-regional communication that is accurate enough to predict multi-regional effects of causal circuit perturbations.

Finally, we evaluated the consistency of \mrlfads across random initializations of the model parameters. We compared \mrlfadsr and \mrlfadsg, each trained using five different random seeds, on simultaneous population recordings from ALM, thalamus, and MRN (see \hyperref[app:exp]{Appendix F2}). Both models inferred consistent effectomes (\Cref{fig:ablation}e), but message content was more consistent across seeds in \mrlfadsr (\Cref{fig:ablation}f, g), further highlighting the benefits of data-constrained communication---particularly for improving the reproducibility of scientific conclusions derived from the model.

\section{Discussion}

Understanding how brain regions interact to support distributed computation requires communication models that can disentangle inter-regional communication from local population dynamics, accounting for region-specific inputs from unrecorded regions. In this work, we identified critical failure modes that limit existing communication models---including misidentification due to manually specified external inputs and mislocalization of neural dynamics due to overly flexible communication architectures. We introduced MR-LFADS, a communication model specifically designed to mitigate such failures through three key design features: (1) automatic inference of region-specific inputs from unobserved sources, (2) data-constrained communication inferred from reconstructed firing rates, and (3) structured regularization that promotes disentangling and prevents inferring inputs from unobserved sources when the same information can be obtained via communication from observed regions. These features discourage the model from learning spurious solutions that explain the data but misrepresent inter-regional interactions. While MR-LFADS is one concrete implementation, the three design principles are more general, and alternative architectures---for example, different formulations of the local-region dynamical systems---may also succeed if they incorporate these same principles.

Using synthetic datasets designed to rigorously test communication models, we demonstrated that \mrlfads outperforms existing approaches in accurately recovering both the structure and content of inter-regional communication. Crucially, ablated model variants indicated that these performance gains stem directly from \mrlfads design features. Applying \mrlfads to real multi-region electrophysiological recordings further validated its utility. \mrlfads inferred inter-regional interactions that accurately predicted brain-wide effects of causal perturbations, despite these perturbations being absent during training. In this setting, \mrlfads models were also more reproducible across random model initializations compared to a model variant that removed data-constraints on inferred communication.

Despite its advantages, \mrlfads has potential limitations. \mrlfads can be sensitive to hyperparameter (HP) settings, mirroring a known limitation of SR-LFADS \cite{keshtkaran2022large}. Thus, significant computational resources might be required to adequately optimize HPs. As in SR-LFADS, HPs specifying the prior distributions (\Cref{eq:prior}) can shape the representations of inferred latent variables, which can in turn shape the inferred population dynamics. Although inferred inputs have been shown to reflect the presence, identity, and timing of external inputs \citep{pandarinath2018inferring}, future work is needed to interpret the representations of those signals \cite{sedler2023expressive, versteeg2024expressive}. The data-constrained communication we propose in this work might mitigate this representational sensitivity in the context of inferred communication.

Another set of \mrlfads HPs (the $\beta$'s from \Cref{sec:mrlfads}) control the model's preference to infer communication rather than inputs from unobserved regions whenever possible. While the settings we chose were effective in this study, different choices may be needed in other scenarios---especially when an unrecorded region directly drives multiple recorded regions. This classical correlation–causation confound cannot be resolved from passively observed neural activity alone. While \mrlfads recovers more causal structure than existing methods, its greatest value may be in generating hypotheses about communication and motivating targeted causal circuit perturbations to test them.

\section*{Code Availability}

A PyTorch implementation of MR-LFADS is available at  \url{https://github.com/golub-lab/MR-LFADS}.

\section*{Acknowledgments} 

This work was supported by NIH award T32-MH132518 (JS), the Paul G. Allen Foundation (MDG), and NIH award R00-MH121533 (MDG). We thank O. Karniol-Tambour for valuable discussion and support with \mrsds \cite{karniol2024modeling}. We are grateful to S. Chen, N. Steinmetz, E. Shea-Brown, and J.N. Kutz for their insights, and to H. Gurnani, T. Kim, and J. Pemberton for thoughtful manuscript feedback. We are also grateful to D. Sussillo, K. Shenoy, W. Newsome, and C. Pandarinath for many years of guidance and inspiration.

\section*{Impact Statement}

This work aims to advance the fields of machine learning and neuroscience, with potential societal impacts including the development of neuroengineering technologies and treatments for neurological injuries, diseases, and neuropsychiatric conditions.

\nocite{raposo2014category}

\bibliography{main}

@article{duncker2021dynamics,
  title={Dynamics on the manifold: Identifying computational dynamical activity from neural population recordings},
  author={Duncker, Lea and Sahani, Maneesh},
  journal={Current Opinion in Neurobiology},
  volume={70},
  pages={163--170},
  year={2021},
  publisher={Elsevier}
}

@article{kass2023identification,
  title={Identification of interacting neural populations: methods and statistical considerations},
  author={Kass, Robert E and Bong, Heejong and Olarinre, Motolani and Xin, Qi and Urban, Konrad N},
  journal={Journal of Neurophysiology},
  volume={130},
  number={3},
  pages={475--496},
  year={2023},
  publisher={American Physiological Society Rockville, MD}
}

@article{keeley2020modeling,
  title={Modeling statistical dependencies in multi-region spike train data},
  author={Keeley, Stephen L and Zoltowski, David M and Aoi, Mikio C and Pillow, Jonathan W},
  journal={Current Opinion in Neurobiology},
  volume={65},
  pages={194--202},
  year={2020},
  publisher={Elsevier}
}

@article{semedo2020statistical,
  title={Statistical methods for dissecting interactions between brain areas},
  author={Semedo, Jo{\~a}o D and Gokcen, Evren and Machens, Christian K and Kohn, Adam and Yu, Byron M},
  journal={Current Opinion in Neurobiology},
  volume={65},
  pages={59--69},
  year={2020},
  publisher={Elsevier}
}

@article{perich2020rethinking,
  title={Rethinking brain-wide interactions through multi-region ‘network of networks’ models},
  author={Perich, Matthew G and Rajan, Kanaka},
  journal={Current Opinion in Neurobiology},
  volume={65},
  pages={146--151},
  year={2020},
  publisher={Elsevier}
}

@article{jia2022multi,
  title={Multi-regional module-based signal transmission in mouse visual cortex},
  author={Jia, Xiaoxuan and Siegle, Joshua H and Durand, S{\'e}verine and Heller, Greggory and Ramirez, Tamina K and Koch, Christof and Olsen, Shawn R},
  journal={Neuron},
  volume={110},
  number={9},
  pages={1585--1598},
  year={2022},
  publisher={Elsevier}
}

@article{siegle2021survey,
  title={Survey of spiking in the mouse visual system reveals functional hierarchy},
  author={Siegle, Joshua H and Jia, Xiaoxuan and Durand, S{\'e}verine and Gale, Sam and Bennett, Corbett and Graddis, Nile and Heller, Greggory and Ramirez, Tamina K and Choi, Hannah and Luviano, Jennifer A and others},
  journal={Nature},
  volume={592},
  number={7852},
  pages={86--92},
  year={2021},
  publisher={Nature Publishing Group UK London}
}

@article{bennett2024shield,
  title={SHIELD: Skull-shaped hemispheric implants enabling large-scale electrophysiology datasets in the mouse brain},
  author={Bennett, Corbett and Ouellette, Ben and Ramirez, Tamina K and Cahoon, Alex and Cabasco, Hannah and Browning, Yoni and Lakunina, Anna and Lynch, Galen F and McBride, Ethan G and Belski, Hannah and others},
  journal={Neuron},
  volume={112},
  number={17},
  pages={2869--2885},
  year={2024},
  publisher={Elsevier}
}

@article{macdowell2025multiplexed,
  title={Multiplexed subspaces route neural activity across brain-wide networks},
  author={MacDowell, Camden J and Libby, Alexandra and Jahn, Caroline I and Tafazoli, Sina and Ardalan, Adel and Buschman, Timothy J},
  journal={Nature Communications},
  volume={16},
  number={1},
  pages={3359},
  year={2025},
  publisher={Nature Publishing Group UK London}
}

@article{steinmetz2019distributed,
  title={Distributed coding of choice, action and engagement across the mouse brain},
  author={Steinmetz, Nicholas A and Zatka-Haas, Peter and Carandini, Matteo and Harris, Kenneth D},
  journal={Nature},
  volume={576},
  number={7786},
  pages={266--273},
  year={2019},
  publisher={Nature Publishing Group UK London}
}

@article{sofroniew2016large,
  title={A large field of view two-photon mesoscope with subcellular resolution for in vivo imaging},
  author={Sofroniew, Nicholas James and Flickinger, Daniel and King, Jonathan and Svoboda, Karel},
  journal={eLife},
  volume={5},
  pages={e14472},
  year={2016},
  publisher={eLife Sciences Publications, Ltd}
}

@article{song2017volumetric,
  title={Volumetric two-photon imaging of neurons using stereoscopy ({vTwINS})},
  author={Song, Alexander and Charles, Adam S and Koay, Sue Ann and Gauthier, Jeff L and Thiberge, Stephan Y and Pillow, Jonathan W and Tank, David W},
  journal={Nature Methods},
  volume={14},
  number={4},
  pages={420--426},
  year={2017},
  publisher={Nature Publishing Group US New York}
}

@article{musall2019single,
  title={Single-trial neural dynamics are dominated by richly varied movements},
  author={Musall, Simon and Kaufman, Matthew T and Juavinett, Ashley L and Gluf, Steven and Churchland, Anne K},
  journal={Nature Neuroscience},
  volume={22},
  number={10},
  pages={1677--1686},
  year={2019},
  publisher={Nature Publishing Group US New York}
}

@article{allen2019thirst,
  title={Thirst regulates motivated behavior through modulation of brainwide neural population dynamics},
  author={Allen, William E and Chen, Michael Z and Pichamoorthy, Nandini and Tien, Rebecca H and Pachitariu, Marius and Luo, Liqun and Deisseroth, Karl},
  journal={Science},
  volume={364},
  number={6437},
  pages={eaav3932},
  year={2019},
  publisher={American Association for the Advancement of Science}
}

@article{makino2017transformation,
  title={Transformation of cortex-wide emergent properties during motor learning},
  author={Makino, Hiroshi and Ren, Chi and Liu, Haixin and Kim, An Na and Kondapaneni, Neehar and Liu, Xin and Kuzum, Duygu and Komiyama, Takaki},
  journal={Neuron},
  volume={94},
  number={4},
  pages={880--890},
  year={2017},
  publisher={Elsevier}
}

@article{allen2017global,
  title={Global representations of goal-directed behavior in distinct cell types of mouse neocortex},
  author={Allen, William E and Kauvar, Isaac V and Chen, Michael Z and Richman, Ethan B and Yang, Samuel J and Chan, Ken and Gradinaru, Viviana and Deverman, Benjamin E and Luo, Liqun and Deisseroth, Karl},
  journal={Neuron},
  volume={94},
  number={4},
  pages={891--907},
  year={2017},
  publisher={Elsevier}
}

@article{gilad2018behavioral,
  title={Behavioral strategy determines frontal or posterior location of short-term memory in neocortex},
  author={Gilad, Ariel and Gallero-Salas, Yasir and Groos, Dominik and Helmchen, Fritjof},
  journal={Neuron},
  volume={99},
  number={4},
  pages={814--828},
  year={2018},
  publisher={Elsevier}
}

@article{stringer2019spontaneous,
  title={Spontaneous behaviors drive multidimensional, brainwide activity},
  author={Stringer, Carsen and Pachitariu, Marius and Steinmetz, Nicholas and Reddy, Charu Bai and Carandini, Matteo and Harris, Kenneth D},
  journal={Science},
  volume={364},
  number={6437},
  pages={eaav7893},
  year={2019},
  publisher={American Association for the Advancement of Science}
}

@article{chen2024brain,
  title={Brain-wide neural activity underlying memory-guided movement},
  author={Chen, Susu and Liu, Yi and Wang, Ziyue Aiden and Colonell, Jennifer and Liu, Liu D and Hou, Han and Tien, Nai-Wen and Wang, Tim and Harris, Timothy and Druckmann, Shaul and others},
  journal={Cell},
  volume={187},
  number={3},
  pages={676--691},
  year={2024},
  publisher={Elsevier}
}

@article{international2023brain,
  title={A brain-wide map of neural activity during complex behaviour},
  author={{IBL} and Benson, Brandon and Benson, Julius and Birman, Daniel and Bonacchi, Niccolo and Bougrova, Kc{\'e}nia and Bruijns, Sebastian A and Carandini, Matteo and Catarino, Joana A and Chapuis, Gaelle A and others},
  journal={bioRxiv preprint},
  year={2023},
  publisher={Cold Spring Harbor Laboratory}
}

@article{kang2020approaches,
  title={Approaches to inferring multi-regional interactions from simultaneous population recordings},
  author={Kang, Byungwoo and Druckmann, Shaul},
  journal={Current Opinion in Neurobiology},
  volume={65},
  pages={108--119},
  year={2020},
  publisher={Elsevier}
}

@article{biswas2020theoretical,
  title={Theoretical principles for illuminating sensorimotor processing with brain-wide neuronal recordings},
  author={Biswas, Tirthabir and Bishop, William E and Fitzgerald, James E},
  journal={Current Opinion in Neurobiology},
  volume={65},
  pages={138--145},
  year={2020},
  publisher={Elsevier}
}

@article{pandarinath2018inferring,
  title={Inferring single-trial neural population dynamics using sequential auto-encoders},
  author={Pandarinath, Chethan and O’Shea, Daniel J and Collins, Jasmine and Jozefowicz, Rafal and Stavisky, Sergey D and Kao, Jonathan C and Trautmann, Eric M and Kaufman, Matthew T and Ryu, Stephen I and Hochberg, Leigh R and others},
  journal={Nature Methods},
  volume={15},
  number={10},
  pages={805--815},
  year={2018},
  publisher={Nature Publishing Group US New York}
}

@article{keshtkaran2022large,
  title={A large-scale neural network training framework for generalized estimation of single-trial population dynamics},
  author={Keshtkaran, Mohammad Reza and Sedler, Andrew R and Chowdhury, Raeed H and Tandon, Raghav and Basrai, Diya and Nguyen, Sarah L and Sohn, Hansem and Jazayeri, Mehrdad and Miller, Lee E and Pandarinath, Chethan},
  journal={Nature Methods},
  volume={19},
  number={12},
  pages={1572--1577},
  year={2022},
  publisher={Nature Publishing Group US New York}
}

@article{zhu2022deep,
  title={A deep learning framework for inference of single-trial neural population dynamics from calcium imaging with subframe temporal resolution},
  author={Zhu, Feng and Grier, Harrison A and Tandon, Raghav and Cai, Changjia and Agarwal, Anjali and Giovannucci, Andrea and Kaufman, Matthew T and Pandarinath, Chethan},
  journal={Nature Neuroscience},
  volume={25},
  number={12},
  pages={1724--1734},
  year={2022},
  publisher={Nature Publishing Group US New York}
}

@article{sedler2023expressive,
  title={Expressive architectures enhance interpretability of dynamics-based neural population models},
  author={Sedler, Andrew R and Versteeg, Christopher and Pandarinath, Chethan},
  journal={Neurons, Behavior, Data Analysis, and Theory},
  volume={2023},
  year={2023},
  publisher={NIH Public Access}
}

@InProceedings{versteeg2024expressive,
  title = 	 {Expressive dynamics models with nonlinear injective readouts enable reliable recovery of latent features from neural activity},
  author =       {Versteeg, Christopher and Sedler, Andrew R. and McCart, Jonathan D. and Pandarinath, Chethan},
  booktitle = 	 {Proceedings of the 2nd NeurIPS Workshop on Symmetry and Geometry in Neural Representations},
  pages = 	 {255--278},
  year = 	 {2024},
  volume = 	 {228},
  series = 	 {Proceedings of Machine Learning Research},
  publisher =    {PMLR},
}

@article{durstewitz2023reconstructing,
  title={Reconstructing computational system dynamics from neural data with recurrent neural networks},
  author={Durstewitz, Daniel and Koppe, Georgia and Thurm, Max Ingo},
  journal={Nature Reviews Neuroscience},
  volume={24},
  number={11},
  pages={693--710},
  year={2023},
  publisher={Nature Publishing Group UK London}
}

@article{linderman2016recurrent,
  title={Recurrent switching linear dynamical systems},
  author={Linderman, Scott W and Miller, Andrew C and Adams, Ryan P and Blei, David M and Paninski, Liam and Johnson, Matthew J},
  journal={arXiv preprint},
  volume={1610.08466},
  year={2016}
}

@article{glaser2020recurrent,
  title={Recurrent switching dynamical systems models for multiple interacting neural populations},
  author={Glaser, Joshua and Whiteway, Matthew and Cunningham, John P and Paninski, Liam and Linderman, Scott},
  journal={Advances in Neural Information Processing Systems},
  volume={33},
  pages={14867--14878},
  year={2020}
}

@inproceedings{karniol2024modeling,
  title={Modeling state-dependent communication between brain regions with switching nonlinear dynamical systems},
  author={Karniol-Tambour, Orren and Zoltowski, David M and Diamanti, E Mika and Pinto, Lucas and Brody, Carlos D and Tank, David W and Pillow, Jonathan W},
  booktitle={The Twelfth International Conference on Learning Representations},
  year={2024}
}

@article{perich2020inferring,
  title={Inferring brain-wide interactions using data-constrained recurrent neural network models},
  author={Perich, Matthew G and Arlt, Charlotte and Soares, Sofia and Young, Megan E and Mosher, Clayton P and Minxha, Juri and Carter, Eugene and Rutishauser, Ueli and Rudebeck, Peter H and Harvey, Christopher D and others},
  journal={bioRxiv preprint},
  year={2020},
  publisher={Cold Spring Harbor Laboratory}
}

@article{kingma2013auto,
      title={Auto-Encoding Variational Bayes}, 
      author={Diederik P Kingma and Max Welling},
      journal={arXiv preprint},
      volume={1312.6114},
      year={2013},
}

@inproceedings{higgins2017beta,
  title={beta-VAE: Learning Basic Visual Concepts with a Constrained Variational Framework},
  author={Higgins, Irina and Matthey, Loic and Pal, Arka and Burgess, Christopher and Glorot, Xavier and Botvinick, Matthew and Mohamed, Shakir and Lerchner, Alexander},
  booktitle={International Conference on Learning Representations},
  year={2017}
}

@article{semedo2019cortical,
  title={Cortical areas interact through a communication subspace},
  author={Semedo, Jo{\~a}o D and Zandvakili, Amin and Machens, Christian K and Yu, Byron M and Kohn, Adam},
  journal={Neuron},
  volume={102},
  number={1},
  pages={249--259},
  year={2019},
  publisher={Elsevier}
}

@article{gokcen2022disentangling,
  title={Disentangling the flow of signals between populations of neurons},
  author={Gokcen, Evren and Jasper, Anna I and Semedo, Jo{\~a}o D and Zandvakili, Amin and Kohn, Adam and Machens, Christian K and Yu, Byron M},
  journal={Nature Computational Science},
  volume={2},
  number={8},
  pages={512--525},
  year={2022},
  publisher={Nature Publishing Group US New York}
}

@article{gokcen2024uncovering,
  title={Uncovering motifs of concurrent signaling across multiple neuronal populations},
  author={Gokcen, Evren and Jasper, Anna and Xu, Alison and Kohn, Adam and Machens, Christian K and Yu, Byron M},
  journal={Advances in Neural Information Processing Systems},
  volume={36},
  year={2024}
}

@article{yu2008gaussian,
  title={Gaussian-process factor analysis for low-dimensional single-trial analysis of neural population activity},
  author={Yu, Byron M and Cunningham, John P and Santhanam, Gopal and Ryu, Stephen and Shenoy, Krishna V and Sahani, Maneesh},
  journal={Advances in Neural Information Processing Systems},
  volume={21},
  year={2008}
}

@article{miller2024cognitive,
  title={Cognitive model discovery via disentangled RNNs},
  author={Miller, Kevin and Eckstein, Maria and Botvinick, Matt and Kurth-Nelson, Zeb},
  journal={Advances in Neural Information Processing Systems},
  volume={36},
  year={2024}
}

@article{kaufman2014cortical,
  title={Cortical activity in the null space: permitting preparation without movement},
  author={Kaufman, Matthew T and Churchland, Mark M and Ryu, Stephen I and Shenoy, Krishna V},
  journal={Nature Neuroscience},
  volume={17},
  number={3},
  pages={440--448},
  year={2014},
  publisher={Nature Publishing Group US New York}
}

@article{ruff2019simultaneous,
  title={Simultaneous multi-area recordings suggest that attention improves performance by reshaping stimulus representations},
  author={Ruff, Douglas A and Cohen, Marlene R},
  journal={Nature Neuroscience},
  volume={22},
  number={10},
  pages={1669--1676},
  year={2019},
  publisher={Nature Publishing Group US New York}
}

@article{veuthey2020single,
  title={Single-trial cross-area neural population dynamics during long-term skill learning},
  author={Veuthey, TL and Derosier, K and Kondapavulur, S and Ganguly, K},
  journal={Nature Communications},
  volume={11},
  number={1},
  pages={4057},
  year={2020},
  publisher={Nature Publishing Group UK London}
}

@article{perich2018neural,
  title={A neural population mechanism for rapid learning},
  author={Perich, Matthew G and Gallego, Juan A and Miller, Lee E},
  journal={Neuron},
  volume={100},
  number={4},
  pages={964--976},
  year={2018},
  publisher={Elsevier}
}

@article{yang2019task,
  title={Task representations in neural networks trained to perform many cognitive tasks},
  author={Yang, Guangyu Robert and Joglekar, Madhura R and Song, H Francis and Newsome, William T and Wang, Xiao-Jing},
  journal={Nature Neuroscience},
  volume={22},
  number={2},
  pages={297--306},
  year={2019},
  publisher={Nature Publishing Group US New York}
}

@article{goris2014partitioning,
  title={Partitioning neuronal variability},
  author={Goris, Robbe LT and Movshon, J Anthony and Simoncelli, Eero P},
  journal={Nature Neuroscience},
  volume={17},
  number={6},
  pages={858--865},
  year={2014},
  publisher={Nature Publishing Group US New York}
}

@article{mante2013context,
  title={Context-dependent computation by recurrent dynamics in prefrontal cortex},
  author={Mante, Valerio and Sussillo, David and Shenoy, Krishna V and Newsome, William T},
  journal={Nature},
  volume={503},
  number={7474},
  pages={78--84},
  year={2013},
  publisher={Nature Publishing Group UK London}
}

@article{watanabe2023tree,
  title={Tree-structured parzen estimator: Understanding its algorithm components and their roles for better empirical performance},
  author={Watanabe, Shuhei},
  journal={arXiv preprint arXiv:2304.11127},
  year={2023}
}

@inproceedings{optuna_2019,
    title={Optuna: A Next-generation Hyperparameter Optimization Framework},
    author={Akiba, Takuya and Sano, Shotaro and Yanase, Toshihiko and Ohta, Takeru and Koyama, Masanori},
    booktitle={Proceedings of the 25th {ACM} {SIGKDD} International Conference on Knowledge Discovery and Data Mining},
    year={2019}
}

@article{liaw2018tune,
    title={Tune: A Research Platform for Distributed Model Selection and Training},
    author={Liaw, Richard and Liang, Eric and Nishihara, Robert
            and Moritz, Philipp and Gonzalez, Joseph E and Stoica, Ion},
    journal={arXiv preprint arXiv:1807.05118},
    year={2018}
}

@article{shenoy2013cortical,
  title={Cortical control of arm movements: a dynamical systems perspective},
  author={Shenoy, Krishna V and Sahani, Maneesh and Churchland, Mark M},
  journal={Annual Review of Neuroscience},
  volume={36},
  number={1},
  pages={337--359},
  year={2013},
  publisher={Annual Reviews}
}

@article{vyas2020computation,
  title={Computation through neural population dynamics},
  author={Vyas, Saurabh and Golub, Matthew D and Sussillo, David and Shenoy, Krishna V},
  journal={Annual Review of Neuroscience},
  volume={43},
  number={1},
  pages={249--275},
  year={2020},
  publisher={Annual Reviews}
}

@article{pospisil2024fly,
  title={The fly connectome reveals a path to the effectome},
  author={Pospisil, Dean A and Aragon, Max J and Dorkenwald, Sven and Matsliah, Arie and Sterling, Amy R and Schlegel, Philipp and Yu, Szi-chieh and McKellar, Claire E and Costa, Marta and Eichler, Katharina and others},
  journal={Nature},
  volume={634},
  number={8032},
  pages={201--209},
  year={2024},
  publisher={Nature Publishing Group UK London}
}

@article{dai2018connections,
  title={Connections with robust PCA and the role of emergent sparsity in variational autoencoder models},
  author={Dai, Bin and Wang, Yu and Aston, John and Hua, Gang and Wipf, David},
  journal={Journal of Machine Learning Research},
  volume={19},
  number={41},
  pages={1--42},
  year={2018}
}

@article{raposo2014category,
  title={A category-free neural population supports evolving demands during decision-making},
  author={Raposo, David and Kaufman, Matthew T and Churchland, Anne K},
  journal={Nature neuroscience},
  volume={17},
  number={12},
  pages={1784--1792},
  year={2014},
  publisher={Nature Publishing Group US New York}
}
\bibliographystyle{icml2025}



\appendix
\onecolumn

\renewcommand{\thefigure}{S\arabic{figure}}
\renewcommand{\thetable}{S\arabic{table}}
\setcounter{figure}{0}
\setcounter{table}{0}

\section{Models}

\subsection{Multi-Region LFADS (MR-LFADS)}
\label{app:mrlfads}

\textbf{Reconstruction validation.} During training, we held out $10\%$ of the neurons in each real or synthetic brain region for validation. After training \mrlfads, we fit a separate linear decoder \textit{post hoc} on the training trials, regressing the inferred factors onto the held-out neurons’ activity. Since the decoder is not trained end-to-end with \mrlfads, held-out neurons do not influence model fitting. Goodness of fit is reported as the $R^2$ between predicted and observed held-out-neuron activity on validation trials.

\textbf{KL penalty.} A critical hyperparameter during training is the scale of the KL penalty. The KL penalty coefficient for the inferred inputs, $\klweight{u}$, is always set higher than that for communication, $\klweight{m}$, an implicit assumption that encourages the network to prioritize learning information via communication channels whenever possible. The timing of when KL penalties are introduced also impacts results, though to a lesser extent. The schedule we found to work well begins with an initial stage where no KL penalty is applied, allowing the model to overfit to the data. Next, the penalty for inferred inputs is introduced, discouraging the model's reliance on these inputs. Finally, the penalty for communication is added, limiting the model from learning excessive information through communication channels.

\textbf{Weight regularization.} We apply light $L_2$ regularization to all GRU network recurrent weights.

\textbf{Hyperparameters.} \Cref{tab:hps} summarizes key hyperparameters used in the \mrlfads models. Overall, we find that KL coefficients have the most impact on held-out neuron loss, $\scos$, and $R^2$ scores for messages compared to other hyperparameters. SR-LFADS was originally described with a factor layer that is potentially lower dimensional than the number of generator units or modeled neurons. Here, we remove the rank constraint from the generator hidden states to rates by setting \( N_\text{fac} = N_\text{neu} \). Additionally, the inferred input and message channel dimensions only need to exceed the estimated true dimensionality of these quantities, as KL penalties naturally suppress redundant channels by driving their activity to zero, as discussed in \hyperref[app:kl_sparsity]{Appendix E.1}.

\begin{table*}[h!]
    \centering
    \caption{Key hyperparameters for \mrlfads models. Experiment 4 refers to applications to multi-region electrophysiology data.}
    \setlength{\extrarowheight}{2pt}
    \begin{tabular}{|l|c|c|}
        \hline
        Hyperparameter & value & Description \\ \hline\hline
        learning rate &   $\in [10^{-5}, 0.004]$   & \raggedright\arraybackslash Scheduled by PyTorch's \texttt{ReduceLROnPlateau}; initial value: $0.004$ \\ \hline
        
        T    &  190  &  Total time steps used for inferring inferred inputs  \\ \hline

        $\tau$    &  10  &  Total time steps used for inferring the initial condition  \\ \hline

        total epoch & 350 & Total number of epochs \\ \hline\hline

        $\klweight{u}$ & & KL penalty coefficient for $u$; performs search for this hyperparameter \\ \hline
        
        $\klweight{u}$ start epoch  &   50    &    Epoch at which $\klweight{u}$ starts increasing from $0$ \\ \hline

        $\klweight{u}$ increase epoch  &   200    &  Number of epochs for $\klweight{u}$ to reach the maximum value  \\ \hline

        $\klweight{m}$ & & KL penalty coefficient for $m$; performs search for this hyperparameter \\ \hline

        $\klweight{m}$ start epoch  &   150    &   Epoch at which $\klweight{m}$ starts increasing from $0$  \\ \hline

        $\klweight{m}$ increase epoch  &   100    &   Number of epochs for $\klweight{m}$ to reach the maximum value  \\ \hline

        $\klweight{\genstate{0}{}}$ & $\klweight{u}$ &  KL penalty coefficient for $\genstate{0}{}$ \\ \hline
        
        $\Ltwoweight$ & $10^4$ &  L2 penalty coefficient \\ \hline

        $\Ltwoweight$ start epoch  &   0    &  Epoch at which $\Ltwoweight$ starts increasing from $0$   \\ \hline

        $\Ltwoweight$ increase epoch  &   80    &    Number of epochs for $\ltwoweight$ to reach the maximum value \\ \hline\hline

        $N_\text{neu}^i$ & & Number of neurons, (64, 16, 64) for exp 1, 2 and 3 respectively \\ \hline

        $N^i_\text{gen}$ & $2N^i_\text{Neu}$ & Generator size \\ \hline

        $N^i_\text{fac}$ & $N^i_\text{Neu}$ & Factor size \\ \hline

        $N^i_\text{inp}$ & & Inferred input dimension, $(4, 8, 6, 8)$ for exp 1, 2, 3 and 4 respectively \\ \hline

        $N^i_\text{msg}$ & & Inferred message dimension, $(4, 16, 10, 8)$ for exp 1, 2, 3 and 4 respectively \\ \hline
    \end{tabular}
    \label{tab:hps}
\end{table*}

\clearpage 
\textbf{Unidirectional Encoder and Controller.} To ensure that inferred inputs reflect only causal information---so that messages, in turn, are causal and thereby allow a more mechanistic interpretation in MR-LFADS---we modify the original LFADS model so that both the encoder and controller used for input inference are entirely unidirectional:
\begin{equation}\label{eq:uenc}
\begin{aligned}
    \encstate{t}{i} &= \encrnn{u}{i}(e^i_{t-1},\data{t}{i}) \\
    c_t^i &= \conrnn{i}(\constate{t-1}{i},[\encstate{t}{i}; \factors{t-1}{i}])\\
\end{aligned}
\end{equation}
The inferred inputs are then given by:
\begin{equation}
\begin{aligned}
    & q(\infinput{t}{i} \mid \data{1:t}{i}) =  q(\infinput{t}{i}\mid \constate{t}{i}) = \mathcal{N}(\mu_{u,t}^i, \Sigma_{u,t}^i) \\
    & \mu_{u,t}^i = W_{\mu_u}^i(c_t^i)  \quad
    \Sigma_{u,t}^i = \mbox{diag} \Big(
    \exp \big( {W}^i_{\sigma_u}(\constate{t}{i}) \big)
    \Big)
\end{aligned}
\end{equation}

By contrast, the encoder for the initial condition can remain bidirectional, since it operates on data preceding $t=1$ and thus does not violate causality:

\begin{equation}
\begin{aligned}
    \encbwdstate{t}{i} &= \encbwdrnng{i}( \encbwdstate{t+1}{i},\data{t}{i}) \\
    \encfwdstate{t}{i} &= \encfwdrnng{i}( \encfwdstate{t-1}{i},\data{t}{i}) \end{aligned}
\end{equation}

\begin{equation}
\begin{aligned}
    q(\genstate{0}{i} \mid \data{-\tau:0}{i}) &=  q(\genstate{0}{i} \mid [\encbwdstate{-\tau}{i}; \encfwdstate{0}{i}]) = \mathcal{N}(\mu_{\genstate{0}{}}^i, \Sigma_{\genstate{0}{}}^i)\\
    \mu_{\genstate{0}{}}^i &= W_{\mu_{\genstate{0}{}}}^i([\encbwdstate{-\tau}{i}; \encfwdstate{0}{i}])\\
    \Sigma_{\genstate{0}{}}^i &= \text{diag}\Big( \exp \big(W_{\sigma_{\genstate{0}{}}}^i([\encbwdstate{-\tau}{i}; \encfwdstate{0}{i}]) \big) \Big)\\
\end{aligned}
\end{equation}

\subsection{Reduced-Rank Regression}
\label{app:rrr}

The RRR model used in this study is based on the inter-regional communication subspace model of \citet{macdowell2025multiplexed}, which integrates reduced-rank regression with ridge regression. The key difference in our implementation is that we apply the rank constraint separately to each source brain region, allowing us to disentangle the contributions of individual areas.
Specifically, rather than concatenating activity from all regions into a single input matrix governed by a shared rank-constrained weight matrix---which conflates signals across regions---we assign a dedicated weight submatrix to each source region $A^j$, each with its own rank constraint $r^j$.

Therefore, for the communication subspace model, we have:   
\begin{equation}
\begin{aligned}
    & W^{\text{rrr}} = \mathop{\arg\min}_{W} \left\| Y - XW \right\|_F^2 + \Ltwoweight \left\| W \right\|_F^2 
    \ \ \ \ \text{s.t. }\text{rank}(W) = r \\
    & \text{which is equivalent to:}\\
    & W^{\text{ridge}} = \arg\min_W \left\| Y - XW \right\|_F^2 + \Ltwoweight \left\| W \right\|_F^2 \\
    & W^{\text{rrr}} = \arg\min_{W}   \left\| XW^{\text{ridge}} - X W \right\|_F^2 
    \ \ \ \ \text{s.t. }\text{rank}(W) = r\\
    & \text{which is then equivalent to:}\\
    & W^{\text{rrr}} = W^{\text{ridge}} V_rV_r^T, \text{where } U \Sigma V^T = X W^{\text{ridge}}
\end{aligned}
\end{equation}
where $X \in \mathbb{R}^{T \times N_\text{src}}$ represents the activity of all source regions concatenated together, and $Y \in \mathbb{R}^{T \times N_\text{tar}}$ represents the activity of the target region. $W^{\text{ridge}}$ is the weight matrix obtained after applying ridge regression, and $\Ltwoweight$ is the ridge regularization parameter. $W^{\text{rrr}}$ is the reduced-rank regression matrix obtained after ridge regression is applied. In the final step, singular value decomposition (SVD) is applied to $X W^{\text{ridge}}$, where $U \Sigma V^T$ is the decomposition, and $V_r$ corresponds to the top $r$ components of $V$. 

In this version, the $W^{\text{ridge}}$ matrix is divided into chunks corresponding to different regions $A^j$:
\begin{align}
    W^{\text{ridge}} = [W^{\text{ridge},1}; W^{\text{ridge},2}; \ldots; W^{\text{ridge},N}],
\end{align}
where each $W^{\text{ridge},j}$ corresponds to the contribution of source region $A^j$, and $[W^1;...;W^N]$ represents a vertical stack of the matrices. SVD is then applied to each individual $W^{\text{ridge},j}$ matrix:
\begin{align}
    U^j \Sigma^j (V^j)^T = XW^{\text{ridge},j}.
\end{align}
The reduced-rank version of $W^{\text{ridge},j}$ is computed as:
\begin{align}
    W^{\text{rrr},j} = W^{\text{ridge},j} V_{r^j}^j (V_{r^j}^j)^T.
\end{align}
Finally, all $W^{\text{rrr},j}$ matrices are concatenated to form the complete $W^{\text{rrr}}$ matrix. This ensures that the rank reduction applied to each $W^{\text{ridge},j}$ only compresses the information within that specific region’s contribution, preserving the interpretability of the communication pathway from $A^j$ to the target region.

The hyperparameters for this model include $\Ltwoweight$ and a matrix $R \in \mathbb{R}^{N \times N}$, where each element $r^{ij}$ represents the rank associated with the communication from source region $A^j$ to target region $A^i$. Additionally, since time delays may exist between regions---and such delays are explicitly configured in the synthetic datasets---a delay parameter $d^{ij}$ is introduced for each source-target communication channel. 

For fitting the memory and pass-decision networks, to tune the model, we perform an iterative search based on cross-validation performance, optimizing the hyperparameters in the following order: $D = \{d^{ij}: i, j = 1, \ldots, N, i \neq j\}$, $\Ltwoweight$, and $R = \{r^{ij}: i, j = 1, \ldots, N, i \neq j\}$. This process is repeated until the hyperparameter values converge. The final values used are provided in \Cref{tab:rrr}. While the rank values $R$ for both networks did not converge exactly to the true ranks of the messages, they were close. Notably, providing the true number of latents did not necessarily lead to better results. The delay values $D$ were accurately learned for both models.

For fitting the networks in Experiment 3, the true delay is directly provided, and other hyperparameters are iterated in the same order for $10$ epochs. 

To increase the robustness of the RRR model fit, we implemented a bagging approach. For each model, 10 trials were bootstrapped from the training set, with each trial containing 200 time steps. A total of 87 fitted models were averaged to obtain the matrix $W^{\text{rrr}}$. This specific number of models was chosen to ensure that the total number of trials used during training remained consistent with other models.

\begin{table}[h!]
    \centering
    \caption{Hyperparameter search results for RRR models.}
    \setlength{\extrarowheight}{2pt}
    \begin{tabular}{|l|c|c|}
        \hline
        & Memory & Pass-Decision \\ \hline
        $\Ltwoweight$ &   0.055   &   0.01   \\ \hline
        $R$    &  
        $\begin{pmatrix}
             & 12 & 24 \\
            12 &  & 32 \\
            24 & 18 & 
        \end{pmatrix}$
        &  
        $\begin{pmatrix}
             & 1 \\
            6 & 
        \end{pmatrix}$  
        \\ \hline
        $d^{ij},\ i \neq j$      &   2    &        0     \\ \hline
    \end{tabular}
    \label{tab:rrr}
\end{table}

\subsection{Multi-Region Switching Dynamical Systems}
\label{sec:mrsds}

We consider two variants of multi-region switching dynamical system models. The first is \mpsrslds \cite{glaser2020recurrent}, which consists of linear transitions, dynamics, and emissions. The relevant hyperparameters are the number of latent states per region, the number of discrete switching states, amount of $L_1$ and $L_2$ regularization on the weights, and the learning rate. Additionally, we consider \mrsds \cite{karniol2024modeling}, which is an extension that uses nonlinear transitions, dynamics, and emissions. It consists of two components: an inference network and a latent state-space model. The inference network is a transformer that performs the amortized inference of latent variables given observed neural activity. The latent state-space model is composed of a number of networks and functions as a structured prior on the latent variables. Specifically, we consider additive communication and input terms to the latent dynamics of the state-space model. That is, messages from other regions and external inputs affected the latent dynamics via additive terms. Relevant hyperparameters include the number of latent states per region, the number of discrete switching states, and the sizes of each sub-network. 

For both models, we did extensive hyperparameter tuning to find the best model for each of the synthetic datasets and then computed all metrics on a held-out test set. We used the Tree-structured Parzen Estimator (TPE) algorithm \cite{watanabe2023tree} with the Optuna backend \cite{optuna_2019} in Ray Tune \cite{liaw2018tune}. The algorithm fits two Gaussian mixture models (GMMs), one to the set of parameter values associated with the best objective and another to the remaining ones. It chooses new parameters to explore by maximizing the ratio of the likelihood between these two GMMs. As such, it is a search strategy which uses results from prior tested hyperparameters to inform the next choice of hyperparameters to test. We used the TPE algorithm to search over all relevant hyperparameters above. Additionally, for \mrsds, we used dropout for regularization with the default settings in the provided implementation. We also manually picked a good learning rate and number of epochs for training. Finally, we also made use of co-smoothing for evaluation. This holds out a set of neurons from the inference network, and computes the fit on the reconstruction of these held-out neurons.

\subsection{Model Comparisons}

We outline the key design features of \mrlfads variants and existing communication models in \Cref{table:design}. Models are compared across four criteria: (1) region-specific dynamics, (2) unsupervised inferred inputs, (3) data-constrained communication, and (4) structured information bottlenecks. Only \mrlfadsr incorporates all four features. 

\mrlfadss ablates the controller, removing inferred inputs and instead using manually specified external inputs for each region. \mrlfadsf and \mrlfadsg both communicate via latent variables not directly grounded in observed data---using factors and generator states, respectively. 

\mrsds and \mpsrslds include region-specific dynamics but rely on external inputs and latent-variable-based messaging, without any regularization on inputs or communication. RRR infers communication from observable quantities but lacks dynamics and inputs altogether. While it enforces a rank constraint on the communication subspace, this is not equivalent to explicit regularization on messages.

\begin{table*}[h!]
    \centering
    \caption{Design features of \mrlfads variants and existing communication models.}
    \setlength{\extrarowheight}{2pt}
    \begin{tabular}{|c|c|c|c|c|c|}
        \hline
        & \begin{tabular}{@{}c@{}}Region-Specific\\Dynamics\end{tabular} 
        & \begin{tabular}{@{}c@{}}Unsupervised\\ Inferred Inputs\end{tabular}  
        & \begin{tabular}{@{}c@{}}Data-Constrained\\ Communication\end{tabular}   & 
        \begin{tabular}{@{}c@{}}Structured Information\\ Bottlenecks\end{tabular} \\
        \hline
        \mrlfadsr & \color{green}\checkmark & \color{green}\checkmark & \color{green}\checkmark & \color{green}\checkmark\\
        \hline
        \mrlfadss & \color{green}\checkmark & \color{red}$\times$ & \color{green}\checkmark &  \color{red}$\times$ \\
        \hline
        \mrlfadsf & \color{green}\checkmark & \color{green}\checkmark & \color{red}$\times$  & \color{green}\checkmark \\
        \hline
        \mrlfadsg & \color{green}\checkmark & \color{green}\checkmark & \color{red}$\times$  & \color{green}\checkmark \\
        \hline
        \mrsds & \color{green}\checkmark & \color{red}$\times$ & \color{red}$\times$ & \color{red}$\times$ \\
        \hline
        \mpsrslds & \color{green}\checkmark & \color{red}$\times$ & \color{red}$\times$ & \color{red}$\times$ \\
        \hline
        RRR & \color{red}$\times$ & \color{red}$\times$ & \color{green}\checkmark & \color{red}$\times$\\
        \hline
    \end{tabular}
    \label{table:design}
\end{table*}

\section{Evaluation Metrics}

\subsection{Quantifying Effectome Similarity}
\label{app:djs}

In a trained \mrlfads model, we define the inferred effectome to be a matrix of pairwise message norms, $M$, with element $M_{i,j}$ as the average value of $||\mu_{m,t}^{j \rightarrow i}||_2$ (\Cref{eq:msg}) across all trials and timesteps. In Experiments 1-2, we compared model-inferred effectomes to the corresponding ground truth connectivity matrix $M_\text{true}$, consisting of ones and zeros to indicate the presence or absence of a communication channel in the DGN, respectively. In contrast, Experiment 3 features networks in which not all connections are actively used; in this case, we define $M_\text{true}$ analogously to $M$, but computed using ground truth messages $m^{j \rightarrow i}$ instead of inferred messages $\mu_{m,t}^{j \rightarrow i}$. To assess similarity between $M$ and $M_\text{true}$, we flatten these matrices into vectors---$\vec{m}$ and $\vec{m}_\text{true}$---and compute their cosine similarity:
\begin{equation}
\begin{aligned}
\scos = \frac{\langle \vec{m}, \vec{m}_\text{true} \rangle}{\|\vec{m}\|_2 \cdot \|\vec{m}_\text{true}\|_2} \in [0, 1],
\end{aligned}
\end{equation}
where perfect alignment is indicated when $\scos = 1$. 

\begin{figure}[H]
    \centering
    \includegraphics[width=\columnwidth]{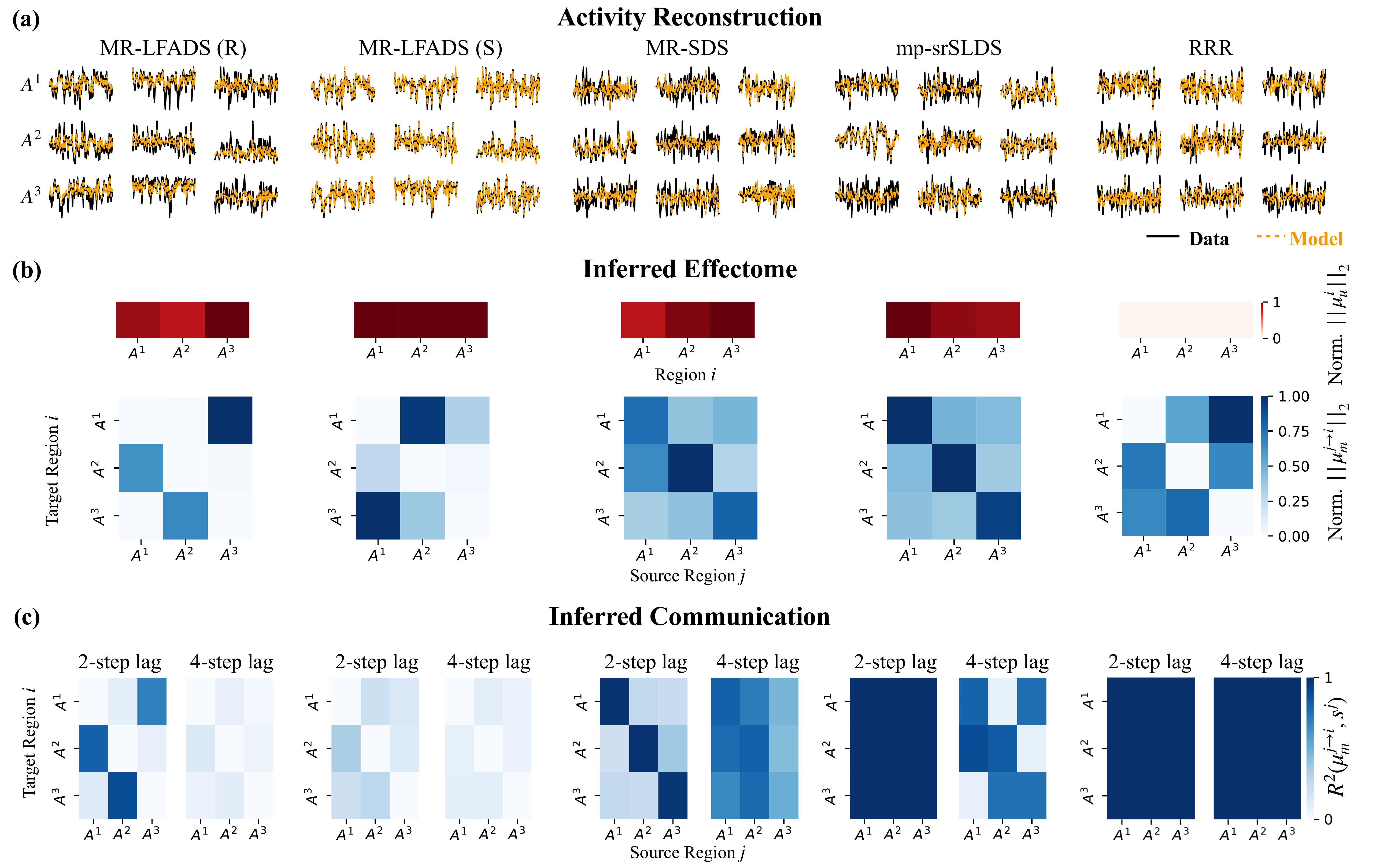}
    \caption{Memory network. (a) Example traces of data reconstruction. \textit{Row:} example neuron from an area, \textit{column:} different trials. (b) Inferred effectomes visualized as heatmaps. \textit{Top:} Inferred inputs, \textit{bottom:} communication. Color intensity represents message norms normalized by the largest input ($\max_{i} ||\mu_u^{i}||_2$) or message norm ($\max_{i,j} ||\mu_m^{j \rightarrow i}||_2$) within each model. These largest inferred inputs are, from left to right: $76$, $175$, $60$, $188$, $0$. The largest messages are: $195$, $22$, $95$, $1431$, $5$. (c). $R^2$ of linear prediction of ground truth messages (with $2$ time step lag on the left, $4$ time step lag on the right) via inferred messages. MR-LFADS(R) results (left) are replicated from \Cref{fig:exp1}e.}
    \label{fig:exp1s}
\end{figure}

\begin{figure}[H]
    \centering
    \includegraphics[width=\columnwidth]{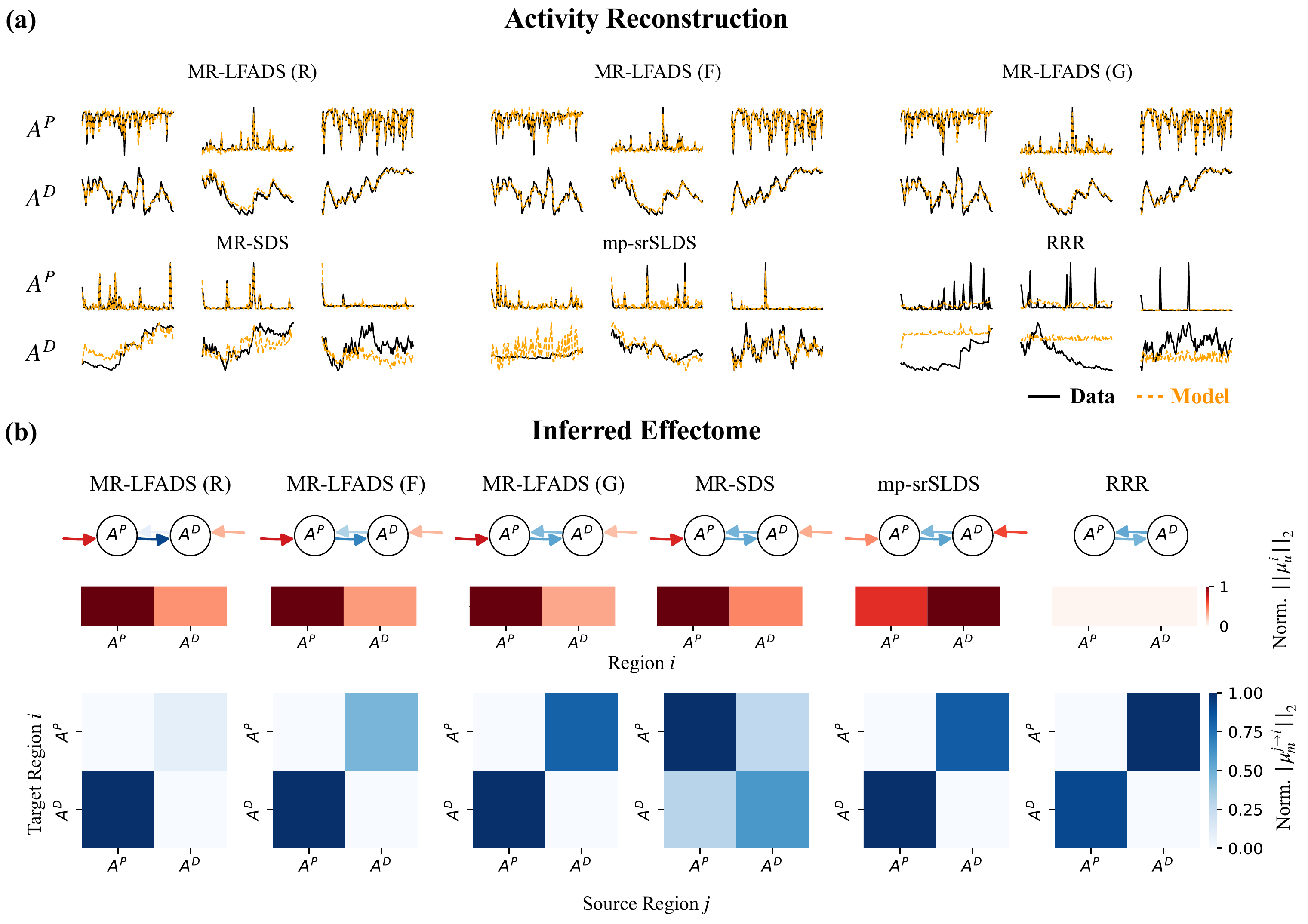}
    \caption{Pass-decision network. (a) Example traces of data reconstruction. \textit{Row:} example neuron from an area, \textit{column:} different trials. (b) Inferred effectomes visualized as circuit diagrams and heatmaps. \textit{Top:} circuit diagram, \textit{middle:} inferred inputs, \textit{bottom:} communication. Color intensity represents message norms normalized by the largest input or message norm within each model.}
    \label{fig:exp2s}
\end{figure}

To visualize an inferred effectome (e.g., \Cref{fig:exp1}c, right), we plot arrows whose color intensity reflects the relative message norm, computed by concatenating all multi-dimensional messages across trial and time, taking the 2-norm, and then normalizing across all channels, with inferred communication and inputs normalized separately. The arrows in \Cref{fig:exp3}e, bottom are further scaled by a sigmoid function for visual contrast. All reported values elsewhere are computed without applying any thresholds or scaling, and heatmap visualizations of the effectome are also provided in \Cref{fig:exp1s}b, \Cref{fig:exp2s}b.

\subsection{Evaluating Information Encoded in Learned Messages} \label{app:msg}

In Experiment 1, we tested whether the inferred inputs and communications encoded information about the past ground truth values, as the network's hidden units activity contains information about past inputs. A correct model should only learn the ground truth inputs or communications. For the results shown in \Cref{fig:exp1}d-e, right, the r-squared values were calculated with a time lag $d$ as $R^2(\mu^{j \rightarrow i}_{m,t}, m^{j \rightarrow i}_{t-d})$.

\section{Synthetic Datasets for System Identification Issues}

Synthetic datasets for networks from Experiments 1-3 have $1024$, $1024$ and $820$ total trials respectively, of which $85\%$ is used for training, and $15\%$ for validation. The length of each trial is $200$ time steps. 

\subsection{Memory Network}
\label{app:memory}

In this synthetic network, each region $A^i$ is modeled as a GRU network with $64$ units that receives a private stimulus $\stim{t}{i} \sim \mathcal{N}(0, \mathbb{I})$ with dimensions $r^i$. To simulate communication channels carrying different amounts of independent information, the dimensions are set as $(r^1, r^2, r^3) = (2, 3, 4)$. Each region has a linear readout $W_{\text{out}}$, and the outputs are required to encode information about the history of all inputs (i.e., stimuli and communication) for up to 5 time steps, enforced using a mean squared error loss. Similarly, the messages transmitted between regions are trained to match $\stim{t-2}{i}$ and are also optimized via a mean squared error loss. Additionally, each region is subjected to dynamic noise $\xi \sim \mathcal{N}(0, 0.01\,\mathbb{I})$, introduced as perturbations to the RNN activity at each time step.

After training the synthetic network, for \mrlfads, we performed a hyperparameter sweep over the KL penalty coefficients for inferred inputs ($\klweight{u} \in \{ 0.01, 0.1, 1, 10\}$) and communication ($\klweight{m} \in \{ 0.001, 0.01\}$). The coefficient pair that resulted in the lowest held-out neuron loss, $(\klweight{u}, \klweight{m}) = (0.1, 0.01)$, was selected. Using these optimized coefficients, we ran the \mrlfads fit across $10$ different seeds, which randomizes model initialization and subsequent sampling of inferred quantities during training, but does not change the allocation of training versus validation data.

Comparing Experiments 1 and 2, it is shown that \mpsrslds and \mrsds activity reconstruction performance underperforms compared to the \mrlfads variants in Experiment~1 (\Cref{fig:exp1}b), but not in Experiment~2 (\Cref{fig:exp2}b). One possible explanation for this discrepancy is the amount of information that the latent variables must encode at each time step. For example, in area $A^1$, the private stimulus is $s^1_t \in \mathbb{R}^2$, and the incoming message from area $A^3$ is $m^{3 \rightarrow 1}_t = s^3_{t-2} \in \mathbb{R}^4$. As a result, the latent representation at time $t$ must capture information spanning 5 time steps and 6 variables in total. Under a standard hyperparameter tuning scheme (Section~\ref{sec:mrsds}), latent variable models like \mpsrslds and \mrsds may struggle to represent all this information accurately.

\subsection{Pass-Decision Network}
\label{app:passdecision}

In this synthetic network, each region is modeled as a GRU network with $16$ units. The stimulus $\stim{t}{}$ is two-dimensional and independently sampled from an exponential distribution with rate parameter of $3$ time steps. We chose a non-Gaussian distribution to test the robustness of MR-LFADS to structural mismatches between the data and the model, as MR-LFADS models the priors and approximate posteriors of inferred inputs and messages as Gaussian.

Each region has a linear readout $W_{\text{out}}$, whose output is required to match its corresponding latent variables ($\stim{t}{}$ for area $A^P$ and $\decision{t}$ for $A^D$). The message sent from $A^P$ to $A^D$ is trained to represent $\stim{t}{}$. Additionally, $A^D$ must encode whether $\decision{t}$ is greater or less than $0$ at all times, mimicking a binary decision-making process.

For the pass-decision network, a low KL penalty for inferred inputs ($\klweight{u} = 0.0075$) was necessary to achieve good held-out neuron loss, while the KL penalty for communication ($\klweight{m} = 0.001$) was set to be approximately one order of magnitude smaller. Using these coefficients, we ran the \mrlfads fit across 10 different seeds.

\section{Randomly Generated Multi-Region Networks}
\label{app:yang}

\begin{table*}[ht!]
    \caption{Task parameters for all families in multi-region data generating networks.}
    \centering
    \begin{tabular}{|l|c|c|c|c|}
        \hline
        \textbf{Task} & \textbf{$\Delta_{\text{offset}}$} & \textbf{$\Delta_{\text{delay}}$} & \textbf{$t_{\text{sacc}}$} & \textbf{$\theta^{\text{resp}}$} \\ \hline\hline
        Go & N/A & $\infty$ & $t_{\text{start}} + \Delta_{\text{dur}}$ & $\theta^{(i)}$ \\ \hline
        Anti-Go & N/A & $\infty$ & $t_{\text{start}} + \Delta_{\text{dur}}$ & $\pi+\theta^{(i)}$ \\ \hline\hline
        Delay-Go & N/A & [30, 50) & $t_{\text{start}} + \Delta_{\text{dur}} + \Delta_{\text{delay}}$ & $\theta^{(i)}$ \\ \hline
        Delay-Anti-Go & N/A & [30, 50) & $t_{\text{start}} + \Delta_{\text{dur}} + \Delta_{\text{delay}}$ & $\pi+\theta^{(i)}$ \\ \hline\hline
        DM1 & 0 & [30, 50) & $t_{\text{start}} + \Delta_{\text{dur}}$ & $\theta^{\text{(1)}}$ \\ \hline
        DM2 & 0 & [30, 50) & $t_{\text{start}} + \Delta_{\text{dur}}$ & $\theta^{\text{(2)}}$ \\ \hline
        MultSen DM & 0 & [30, 50) & $t_{\text{start}} + \Delta_{\text{dur}}$ & $\theta^{\text{(i)}},\ i = \arg\max_i r^{(i)}$ \\ \hline\hline
        Delay-DM1 & [10, 20) & [30, 50) & $t_{\text{start}} + \Delta_{\text{dur}} + \Delta_{\text{offset}}$ & $\theta^{\text{(1)}}$ \\ \hline
        Delay-DM2 & [10, 20) & [30, 50) & $t_{\text{start}} + \Delta_{\text{dur}} + \Delta_{\text{offset}}$ & $\theta^{\text{(2)}}$ \\ \hline
        Delay MultSen DM & [10, 20) & [30, 50) & $t_{\text{start}} + \Delta_{\text{dur}} + \Delta_{\text{offset}}$ & $\theta^{\text{(i)}},\ i = \arg\max_i r^{(i)}$ \\ \hline\hline
        Angle & [10, 20) & [30, 50) & $t_{\text{start}} + \Delta_{\text{dur}} + \Delta_{\text{offset}}$ if $\theta^{\text{(1)}} = \theta^{\text{(2)}}$ & $\theta^{\text{(2)}}$ \\ \hline
        Anti-Angle & [10, 20) & [30, 50) & $t_{\text{start}} + \Delta_{\text{dur}} + \Delta_{\text{offset}}$ if $\theta^{\text{(1)}} = \theta^{\text{(2)}}$ & $\pi + \theta^{\text{(2)}}$ \\ \hline\hline
        Category & [10, 20) & [30, 50) & $t_{\text{start}} + \Delta_{\text{dur}} + \Delta_{\text{offset}}$ if $\text{sign}(\theta^{\text{(1)}}) = \text{sign}(\theta^{\text{(2)}})$ & $\theta^{\text{(2)}}$ \\ \hline
        Anti-Category & [10, 20) & [30, 50) & $t_{\text{start}} + \Delta_{\text{dur}} + \Delta_{\text{offset}}$ if $\text{sign}(\theta^{\text{(1)}}) = \text{sign}(\theta^{\text{(2)}})$ & $\pi + \theta^{\text{(2)}}$ \\ \hline
    \end{tabular}
    \label{tab:tasks}
\end{table*}

We generated a distribution of networks designed to perform computational tasks inspired by \citet{yang2019task}. Each network consists of either 3 or 4 regions. A connection probability $p \in \{0.5, 0.6, 0.7\}$ is specified, and the connectome is randomly drawn. To be considered valid, the connectome must meet two criteria: (1) each region must have at least one input connection and one output connection to ensure no region is redundant, and (2) all regions must be within a maximum distance of 2 steps from the output region. Once a valid connectome is generated, the network is trained on one of the computational tasks. During training, dynamic noise $\xi \sim \mathcal{N}(0, 0.01\,\mathbb{I})$ is applied to all regions, and the only loss is based on whether the output region produces the correct response. Networks that meet the performance thresholds (train accuracy $> 0.8$ and validation accuracy $> 0.6$) are selected as synthetic datasets.

For this case, since we aim to collect a distribution of results, we do not perform a hyperparameter sweep over the KL penalties. Instead, we fit one instance of each communication model to each synthetic dataset using a single random seed. The computational tasks inspired by \citet{yang2019task} are described below.

All trials are 200 time steps in length. Each task receives a fixation input $s_{\text{fix}, t}$, stimuli from two channels $s^{\text{(1)}}_t = (a^{\text{(1)}}, \theta^{\text{(1)}})$ and $s^{\text{(2)}}_t = (a^{\text{(2)}}, \theta^{\text{(2)}})$, and requires a saccade response $r^{\text{sacc}}_t$ and an additional response $r_{\text{resp},t} = (1, \theta_{\text{resp}})$, which differ based on the specified task. For different tasks, stimuli may come from one or both of the channels. Both the stimuli and responses are expressed in polar coordinates, with a resolution of 10 degrees per angle.

The tasks are described in terms of $3$ different families: the go task family, context-dependent decision-making family, and matching family. For all tasks, $t_{\text{start}} \in [30, 50)$ denotes the onset of the first (and sometimes only) stimulus (\Cref{tab:tasks}). The duration of all stimulus pulses in a trial is represented by $\Delta_{\text{dur}}  \in [30, 50)$, and $\Delta_{\text{offset}}$ specifies the offset between the two stimuli, if applicable. The time between the last stimulus offset and the fixation cue offset is given by $\Delta_{\text{delay}}$, where $\Delta_{\text{delay}} = \infty$ indicates that the fixation cue never disappears. The onset of the saccade is denoted as $t_{\text{sacc}}$. Each parameter---$t_{\text{start}}$, $\Delta_{\text{dur}}$, $\Delta_{\text{offset}}$, and $\Delta_{\text{delay}}$---is drawn independently and uniformly from its specified half-open interval $[t_1, t_2)$.

\subsection{Go Task Family}

The common characteristic of tasks in this family is that only one of the stimulus channels contains the signal, which varies between trials. Depending on the specific task, the network must saccade dependently or independently of the fixation cue. The response is required to be either in the direction of the signal pulse, $\theta^{(i)}$, or in the opposite direction, $\pi + \theta^{(i)}$. The individual tasks are summarized in \Cref{tab:tasks}.

\subsection{Context-Dependent Decision-Making Family}

For this family of tasks, stimulus pulses occur in both channels, and the network must report either $\theta^{\text{(1)}}$ or $\theta^{\text{(2)}}$, depending on the specific task type. In some tasks, the pulses occur at different times, requiring the network to maintain memory of the stimuli. The task parameters for this family are summarized in \Cref{tab:tasks}.

\subsection{Matching Family}

In the matching tasks, the network determines whether to saccade based on whether the two stimulus angles “match.” In the “Angle” tasks, the network saccades only if $\theta^{\text{(1)}} = \theta^{\text{(2)}}$ under the given resolution (10 degrees per angle). In the “Category” tasks, the network saccades if $\text{sign}(\theta^{\text{(1)}}) = \text{sign}(\theta^{\text{(2)}})$, meaning both angles are either positive or negative. Task details are provided in \Cref{tab:tasks}. Regardless of whether the angles match, the response is always set to report the angle $\theta^{\text{(2)}}$ (or the opposite of it).

\begin{figure}[ht!]
    \centering
    \includegraphics[width=\columnwidth]{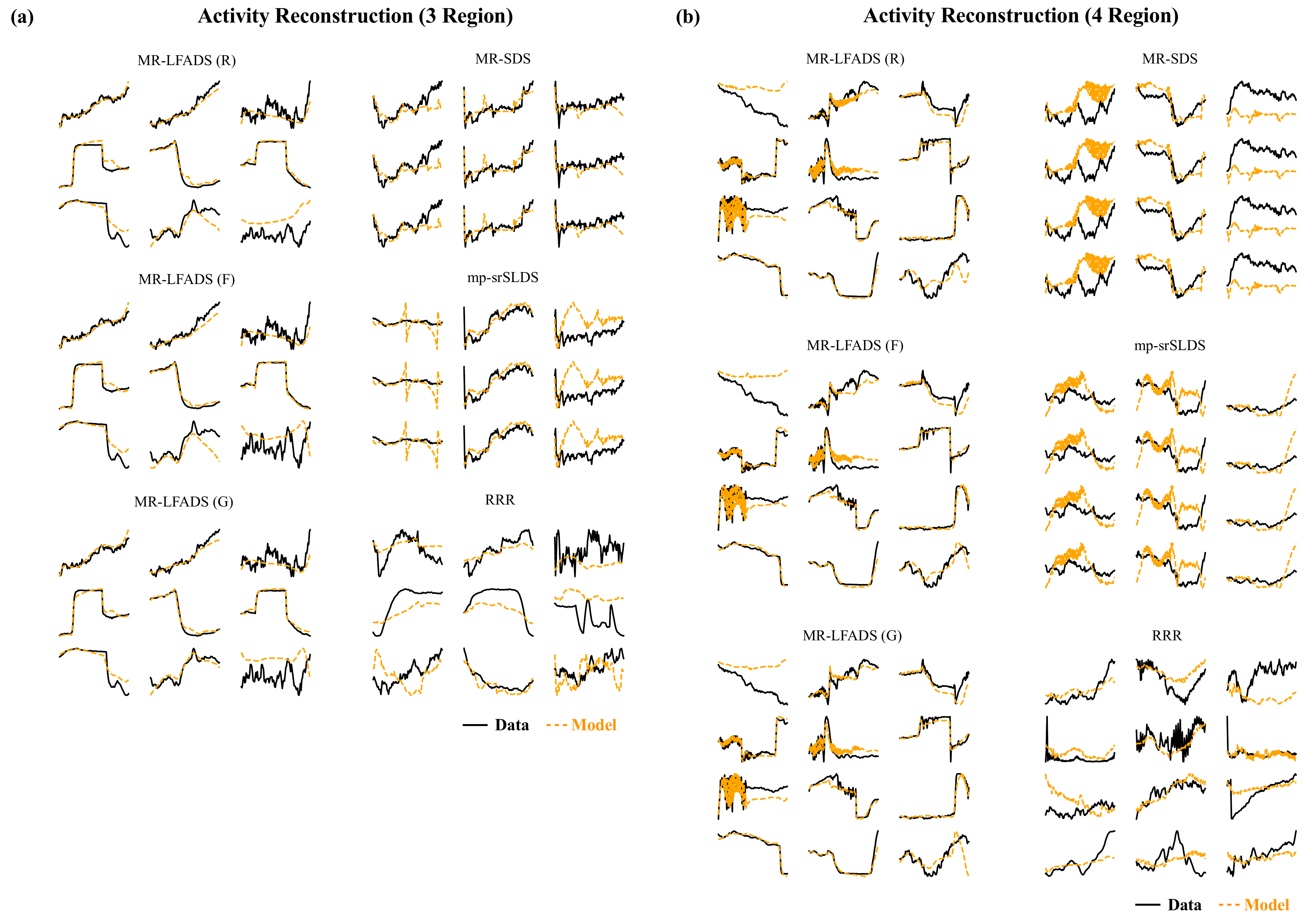}
    \caption{Randomly generated multi-region network: example traces of data reconstruction for networks with median \mrlfadsr performance for 3 regions (a) and 4 regions (b).}
    \label{fig:exp3s1}
\end{figure}

\clearpage
\section{Implications of Constrained Architectural Choices in Restricting Message Content}

\subsection{Input and Message Inference with KL Penalties}
\label{app:kl_sparsity}

\begin{figure}[H]
    \centering
    \includegraphics[width=0.8\columnwidth]{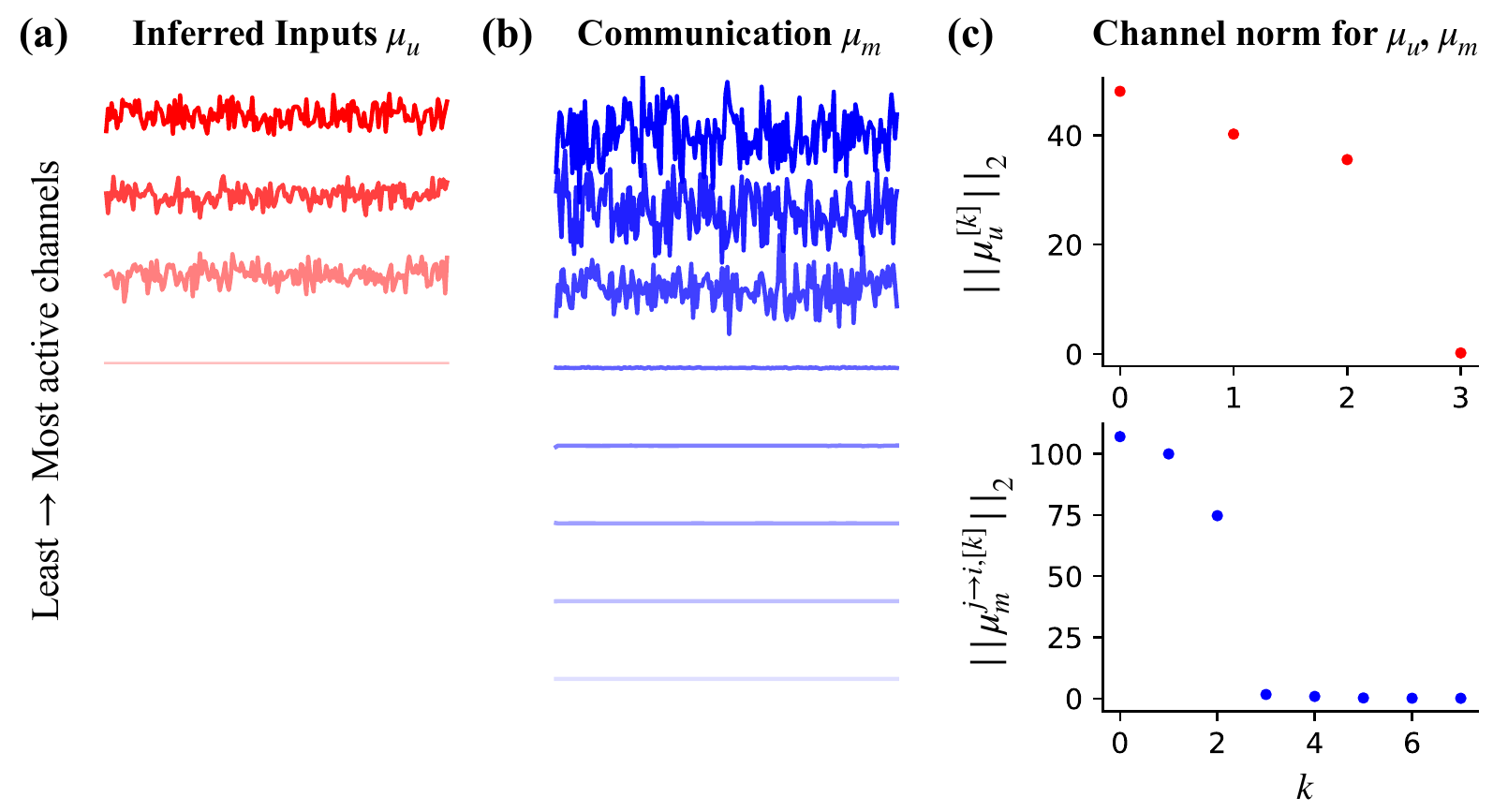}
    \caption{The effect of KL penalties on restricting information across inferred inputs and communication channels in Experiment~1. (a) Inferred input norm over time, $||\mu_{u,t}^{[k]}||_2$, for the most (dark red) and least (light red) active channels, where $k$ denotes channel number. (b) Inferred message norm over time, $||\mu_{m,t}^{j \rightarrow i, [k]}||_2$, for the most (dark blue) and least (light blue) active channels. (c) Inferred input (top) and message (bottom) norm across channels for area $A^1$ in descending order of its scalar norm across trials and time.}
    \label{fig:kl_sparsity}
\end{figure}

KL penalties with standard Gaussian priors in variational autoencoders are known to reduce latent space dimensionality by pruning unnecessary dimensions \cite{dai2018connections, miller2024cognitive}. Consequently, with sufficiently high KL penalty coefficients $(\klweight{u}$, $\klweight{m})$, \mrlfads is incentivized to use only the communication channels essential for data reconstruction. This effect is evident when comparing the most active channel (i.e., the one with the highest input or message norm across trials and time) to the least active ones (\Cref{fig:kl_sparsity}a, b). Examining input and message norms across all channels further confirms that some channels are effectively silenced (\Cref{fig:kl_sparsity}c).

\subsection{Message Inference via Rates versus Factors}
\label{app:sing_val}

\begin{figure}[H]
    \centering
    \includegraphics[width=0.6\columnwidth]{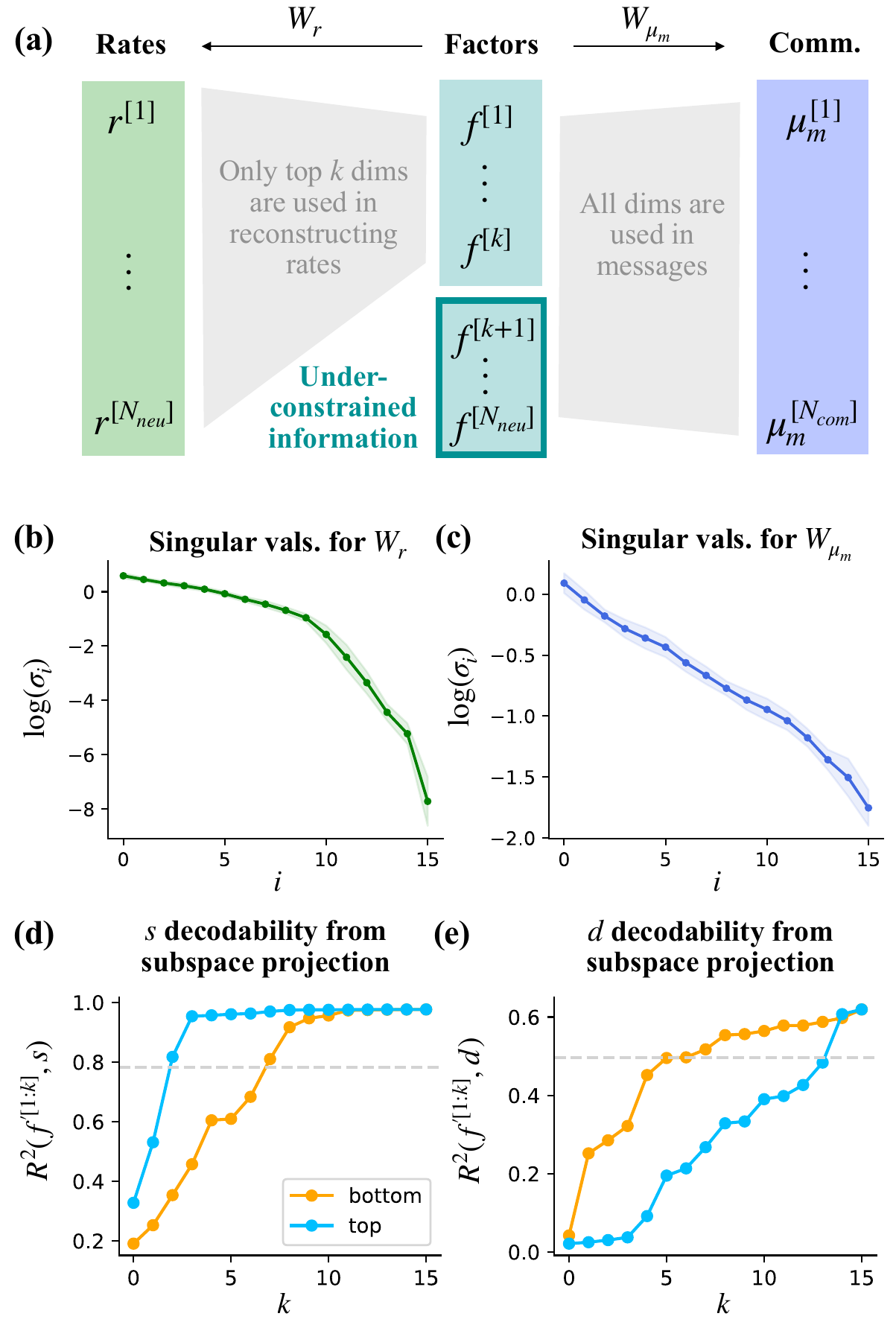}
    \caption{SVD Analysis on \mrlfadsf. (a) Illustration of factors containing unconstrained information. Rates, factors, and communication are all from area P, i.e. $r \equiv r^p$, $f \equiv f^p$, $\mu_m \equiv \mu_m^{P \rightarrow D}$. $k$ is the effective rank of $W_r$. (b) Ranked singular values for $W_r$. (c) Ranked singular values for $W_{\mu_m}$. Shaded regions indicate standard deviation across different random initializations (seeds).
    (d) $R^2$ value for decoding $s$ from $f^P$ projected onto subspaces spanned by top / bottom left singular vectors of $W_r$ of an example seed. (e) Same as (d), but for decoding $d$.}
    \label{fig:sing_val}
\end{figure}

Since \mrlfads factors and reconstructed rates share the same dimensionality and are related by a linear transformation, it may not be immediately clear how message inference from these variables leads to substantial differences. To investigate this, we examine \mrlfadsf trained on the pass-decision synthetic data. We perform SVD on projection matrices (\Cref{fig:sing_val}a) from factors to rates, $W_r$ (\Cref{eq:gaussianparams}), and from factors to communication, $W_{\mu_m}$ (\Cref{eq:msgparams} with $r^j_t$ replaced by $f^j_t$). Our analysis shows that $W_r$ has small singular values, indicating that certain factor dimensions contribute minimally to the reconstructed rates (\Cref{fig:sing_val}b). In contrast, $W_{\mu_m}$ exhibits relatively uniform singular values (within the same order of magnitude), suggesting that all factor dimensions are utilized in communication (\Cref{fig:sing_val}c). This implies that some factor dimensions play a role in message inference while being largely detached from rate reconstruction (\Cref{fig:sing_val}a).

To further investigate, we re-express the factor space of area P using the left singular vectors of $W_r$, denoted $U$. In Experiment 2, \mrlfadsf is shown to encode both the stimulus \( s \) and decision variable \( d \) in its \( m^{P \rightarrow D} \) messages (\Cref{fig:exp2}e). To examine how these variables are distributed across the subspaces of \( U \) and whether this aligns with the under-constrained dimensions identified earlier, we project the factors \( f^P \) onto the subspace spanned by the top \( k \) singular vectors of \( U \), denoted \( U^{[1:k]} \), varying \( k \) from 1 to \( N_\text{msg}^D \). We then compute the $R^2$ values for predicting \( s \) and \( d \) from the projected values \( f' \) (\Cref{fig:sing_val}d, e, cyan lines). We repeat this process in the reverse direction, projecting onto the bottom \( k \) singular vectors instead (\Cref{fig:sing_val}d, e, orange lines). The results reveal a clear separation: decoding accuracy for \( s \) improves more when projecting onto the top singular vectors, whereas decoding accuracy for \( d \) increases more rapidly when projecting onto the bottom singular vectors. This suggests that information not used for rate reconstruction---such as \( d \)---is preferentially encoded in the under-constrained dimensions of the latent space, reinforcing the idea that message inference utilizes latent dimensions beyond those needed for rate prediction.

\section{Application to Multi-Region Electrophysiological Data}
\label{app:exp}

Data analyzed are from mouse ID:440959. For the photoinhibition experiment, we analyzed a session with the following recorded regions:
\begin{itemize}
    \item \textbf{Anterior Lateral Motor cortex (ALM)}: MOs2/3, MOs5, and MOs6 (layer 6)
    \item \textbf{Thalamus (ALM, A)}: VM and VAL
    \item \textbf{Thalamus (Other, O)}: Anterior Ventral (AV) and Lateral Dorsal (LD)
    \item \textbf{Midbrain Reticular Nucleus (MRN)}: MRN
    \item \textbf{Superior Colliculus (SC)}: intermediate gray (SCig), intermediate white (SCiw), optic (SCop), superficial gray (SCsg), and zonal layer (SCzo)
\end{itemize}

For the model consistency experiment, we analyzed a session with the these recorded regions:
\begin{itemize}
    \item \textbf{ALM}: secondary motor cortex, layers 2/3 and 5 (MOs2/3, MOs5)
    \item \textbf{Thalamus}: Ventral Medial (VM) and Ventral Anterior-Lateral (VAL)
    \item \textbf{MRN}: MRN
\end{itemize}

Trials were filtered to include only those with durations between $4.5$ and $5.5$ seconds. Each trial was binned into $500$ time steps, with each bin corresponding to a $10$ ms interval.

\subsection{Comparison of Photoinhibition Effects}

We selected one session of the data involving five brain regions previously implicated in a decision-making task, as identified in \citet{chen2024brain}. For both control and photoinhibition trials, we only used trials from the same condition per comparison---either left-hit or right-hit---where ``hit" indicates making a correct choice, and ``left" or ``right" refers to the correct choice. For photoinhibition trials, we focused on those with perturbations within the delay period, aligning all such trials to the photoinhibition onset. Firing rate $\bar{r}_t$ is smoothed from raw spike counts using a causal exponential filter with rate parameter of $7.1$. For each region, we computed the absolute difference in trial-averaged activity between photoinhibited and control trials, averaged over neurons, to estimate the influence of photoinhibition.

\subsection{Consistency of Model Inference Across Random Seeds}

To ensure a fair comparison between models, we evaluated each using the same hyperparameters and same number of random seeds. We then computed all pairwise similarities between inferred effectomes and messages across seeds to assess the consistency of model inference.

\clearpage

\end{document}